\documentclass[]{aa}  

\usepackage{subcaption}
\usepackage{graphicx}
\usepackage{tabularx}
\usepackage{txfonts}
\usepackage{placeins}
\usepackage{xcolor}
\usepackage{hyperref}
\hypersetup{colorlinks=true,
    linkcolor=blue,citecolor=blue}
\usepackage{comment}

\newcommand{\prodimo}{P{\tiny RO}D{\tiny I}M{\tiny O}\;}

\begin{document}

   
    \title{The influence of accretion bursts on methanol and water in massive young stellar objects}
   
\author{R. Guadarrama
          \inst{1},
          E. I. Vorobyov\inst{1,2}, Ch. Rab\inst{3,4}, M. Güdel\inst{1,5}, A. Caratti o Garatti\inst{6,7}, A. M. Sobolev\inst{8} 
          }
\authorrunning{Guadarrama et al.}   
\institute{Department of Astrophysics, University of Vienna,
              T\"urkenschanzstrasse 17, A-1180 Vienna, Austria \\
             \email{rodrigo.guadarrama@univie.ac.at}
             \and
Research Institute of Physics, Southern Federal University, Rostov-on-Don, 344090 Russia
\and Max-Planck-Institut für extraterrestrische Physik, Giessenbachstrasse 1, 85748 Garching, Germany
\and University Observatory, Faculty of Physics, Ludwig-Maximilians-Universität München, Scheinerstr. 1, 81679 Munich,
Germany
\and Max-Planck-Institut für Astronomie, Königstuhl 17, 691176 Heidelberg, Germany
\and NAF-Osservatorio Astronomico di Capodimonte,   via Moiariello 16, I-80131 Napoli, Italy
\and Dublin Institute for Advanced Studies, School of Cosmic Physics, Astronomy \& Astrophysics Section, 31 Fitzwilliam Place, Dublin 2, Ireland
\and Ural Federal University, 19 Mira Str., 620002 Ekaterinburg, Russia
             }
   
\titlerunning{The influence of accretion bursts in MYSOs}

 
  \abstract
  {The effect of accretion bursts on massive young stellar objects (MYSOs) represents a new research field in the study of young stars and their environment. The impact of such bursts on the disc and envelope has been observed and plays the role of a "smoking gun" providing information about the properties of the burst itself.}   
   {We aim to investigate the impact of an accretion burst on massive disks with different types of envelopes and to study the effects of an accretion burst on the temperature structure and the chemistry of the disk. We focus on water and methanol as chemical species for this paper.}
   {The thermochemical code of \prodimo {(PROtoplanetary DIsk MOdel)} is used to perform simulation of high mass protoplanetary disk models with different types of envelopes under the presence of an accretion burst. The models in question represent different evolutionary stages of protostellar objects. We calculate and show the chemical abundances in three phases of the simulation (pre-burst, burst, and post-burst).}
   {More heavily embedded disks show higher temperatures. The impact of the accretion burst is mainly characterized by the desorption of chemical species present in the disk and envelope from the dust grains to the gas phase. When the post-burst phase starts, the sublimated species freeze out again. The degree of sublimation depends strongly on the type of envelope the disk is embedded in. An accretion burst in more massive envelopes produces stronger desorption of the chemical species. However, our models show that the timescale for the chemistry to reach the pre-burst state is independent of the type of envelope.}
    {The study shows that the disk's temperature increases with a more massive envelope enclosing it. Thus, the chemistry of MYSOs in earlier stages of their evolution reacts stronger to an accretion burst than at later stages where the envelope has lost most of its mass or has been dissipated. The study of the impact of accretion bursts could also provide helpful theoretical context to the observation of methanol masers in massive disks.}
   \keywords{stars:protostars-stars: high mass protoplanetary disks-methods: numerical }

   \maketitle
%

\section{Introduction}
Accretion is a fundamental process in star formation and disk evolution. It provides the growing protostar with mass and angular momentum. On the other hand, accretion feeds the gravitational energy of accreted matter back to the disk and circumstellar environment via accretion luminosity. Although it was thought to be a smooth and continuous process, it is now established that unsteady and episodic accretion is an intrinsic part of forming young stars \citep{1996ARA&A..34..207H,2014prpl.conf..387A,2018ApJ...861..145C,2022arXiv220311257F}.

The study of this phenomenon has mainly focused on low-mass stars for historical reasons since low-mass classical TTauri stars (or pre-main sequence stars) are or become optically {visible} while still accreting and are therefore easy to detect.  {One example of a study where the water snowline of a low-mass star has been observed in the context of an accretion burst is the work of \cite{2016Natur.535..258C}.  {Another example is the work of \cite{2023Natur.615..227T}. Here they report the direct detection of gas-phase water from the disk of V883 Ori and measure the water snowline radius at the midplane.
Other works that have studied the effect of accretion bursts on the chemistry of low-mass objects are \cite{2016ApJ...821...46T} and \cite{2018ApJ...866...46M}. In the first one, the processes of gas-phase formation and recondensation of some complex organic molecules due to sudden evaporation processes during luminosity outbursts are studied. In the latter, they identify gas-phase tracers of the burst activity that could be observed with telescopes like ALMA and NOEMA.}

In contrast, high-mass stars (stars that have $M_* > 8 \times M_{\sun}$) are considered to evolve much faster and to be still deeply embedded in their envelope when reaching the main sequence.  {Their rapid formation and therefore scarcity, made it difficult to know if high-mass stars also experience episodic accretion events exhibited in their formation. However, recent studies have approached this subject and delivered results of observed high-mass bursts.} For example, \cite{2017NatPh..13..276C} reports the first detected disk-mediated accretion burst from the source S255IR NIRS 3. This MYSO has a mass of $\approx$ 20 M$_\sun$  and a pre-burst luminosity of $\approx$ 2.4 $\times 10^4$  L$_\sun$. The discovery was triggered by the detection of class II methanol maser flares (at 6.7 GHz), that are excited by IR radiation from the dusty disk heated by the burst.
Near-infrared spectroscopy shows emission lines of HI and CO typically found in accretion disks of low-mass YSOs, but orders of magnitudes brighter. Together with other results shown in that study \citep{2017NatPh..13..276C}, disk accretion is established as a typical mechanism for all YSOs. Furthermore, monitoring methanol maser flares as a way to identify accretion bursts in MYSOs was confirmed in \cite{2021A&A...646A.161S}.         

There are also studies that use models to analyze the behavior of MYSO bursts. \cite{2015ApJ...805..115V} showed that stars form via a steady low-rate accretion accompanied by short and intensive accretion bursts.
The simulations used by \cite{2017MNRAS.464L..90M} showed that the inward migration of circumstellar gaseous clumps towards the star is the mechanism that triggers the accretion bursts. \cite{2021MNRAS.500.4448M} studied the impact of the mass of the prestellar core and the rotational-to-gravitational energy ratio, on the accretion and the evolution of the protostellar mass. Among the main findings of that study are the following: a) Cores with higher mass and higher rotational-to-gravitational energy ratios are more prone to experience accretion bursts. b) Almost all massive protostars have accretion bursts. c) About 40-60\% of the mass is accreted in bursts. Other theoretical studies \citep{2021A&A...651L...3E} find that disks around MYSOs are much hotter than disks around lower-mass YSOs. This would cause, for example, an extended region of the disk to be susceptible to thermal hydrogen ionization, which could produce instabilities and trigger periodic bursts. Although this would potentially offer burst periodicity signatures, the burst produced by this type of instability would be too long and too low in mass transfer to power recently observed bursts in MYSOs. In their study, \cite{2021A&A...651L...3E} showed that the observed bursts are well explained by the tidal disruption of young gas giant planets.

 {In this study, we focus on the impact of an accretion burst on the chemical abundances of methanol and water. To achieve this, we analyze the time-dependent chemical evolution of H$_2$O and CH$_3$OH before, during, and after the accretion burst.  We also track the evolution of the respective snow regions of these two molecules.  {We consider H$_2$O because of its important role as a tracer for different disk properties. Lines of H$_{2}$O in the IR regime can provide information about the temperature in the disk \cite[]{2013ChRv..113.9016H} and it can also potentially be used as a mass tracer, alternatively to CO \cite[]{2017ApJ...849..130M}.} The position of the water snowline is also relevant for dust growth in the disk, planet formation, and composition. 
The role of CH$_3$OH in the study of protoplanetary disks is also of importance.  {It is a complex organic molecule (COM) and also contributes to the formation of more complex organic molecules.} Additionally, the detection of methanol maser flares has been used in several studies to confirm the occurrence of accretion bursts in MYSOs (\citealt{2017NatPh..13..276C}, \citealt{2019ApJ...881L..39B}, \citealt{2019MNRAS.489.3981M}, and \citealt{2021yCat..51610071H}).}
 
In Section \ref{Method} we give an insight into the model used for this study and the method that was used to simulate an accretion burst. In Section \ref{Results} we present the resulting chemical abundances for the pre-burst phase, the burst phase, and the post-burst phase. Additionally, we study the possible occurrence of methanol masers in our models. In section \ref{Discussion} we discuss our results and possible future studies and in Section \ref{Summary and conclusions} we summarize and present our main findings. 

\section{Method}
\label{Method}
 {Low-mass young stellar objects are traditionally classified by the shape of their spectral energy distribution (SED) in the infrared (IR). Class I objects are considered to be embedded in a protoplanetary disk and envelope. Class II objects are optically visible young stars with a surrounding disk but with a heavily depleted or no remaining envelope. 
However, massive young stellar objects are typically not classified in the same way. In general terms, deeply embedded massive early-stage protostars ($10^3$ yr age) are invisible at NIR and MIR wavelengths.  {More evolved massive protostars ($10^4$ yr age) become visible in the IR.} When reaching the age of $10^5$ yr, MYSOs become optically visible, already well into the main sequence and with the envelope and most of the disk gone. 
In this study, we model three massive young stellar objects. Two are representative of a young embedded stage ($10^4$ yr)  and one represents a rather idealized case of a more evolved object ($10^5$ yr) object. {The latter one is a model that corresponds to the onset of the ZAMS (zero-age main sequence) and we use this model to compare more realistic models with envelopes.}
  Thus, the difference between the models is the level to which the disk is embedded in the envelope. 

  {The cavity opening angle of the envelope with respect to the disk axis and the mass of the envelope are the main elements of the modeling setup.} By changing these two parameters, we intend to represent our two different stages of massive disks.  {We note that this implies a simplification in the evolution of MYSOs as the mass of the disk also decreases and disperses after the envelope has dissipated.} The properties of the respective star should also change depending on the age of the source. However, we assume these simplified models to be able to directly compare the effect of the envelope on the chemistry in the context of a burst.

\begin{table}[]
\caption{Main parameters of the MYSO model.}
    \label{table:1}
    \centering
    \begin{tabular}{l|c|c} 
    \hline
    \hline
    Quantity & Symbol & Value \\
    \hline
    stellar mass & $M_{*}$ & 10.0 $M_{\sun}$ \\ 
    stellar effective temp. & $T_{*}$ & 9045 K \\ 
    stellar luminosity & $L_{*}$ & 3241.0 $L_{\sun}$ \\
    accretion Luminosity & $L_{\mathrm{accr}}$ & 10000 $L_{\sun}$ \\
    accretion Luminosity (burst) & $L_{\mathrm{accr_b}}$ & 100000 $L_{\sun}$ \\
    \hline
    disk gas mass & $M_{\mathrm{disk}}$ & 5.0 $M_{\sun}$ \\
    disk inner radius & $R_{\mathrm{in}}$ & 2.0 au \\
    disk tapering-off radius & $R_{\mathrm{tap}}$ & 300 au \\
    
    reference scale height &{ $H(100\ au)$} & 10 au \\
    flaring power index & $\beta$ & 1.25 \\
    \hline
    outer radius & $R_{\mathrm{out}}$ & 10000 au \\
    \hline 
    dust to gas mass ratio & $\delta$ & 0.01 \\
    min. dust particle radius & $\alpha_{\mathrm{min}}$ & 0.005 $\mu$m \\
    max. dust particle radius & $\alpha_{\mathrm{max}}$ & 1000 $\mu$m \\
    dust size power ind. & $\alpha_{\mathrm{pow}}$      &  3.5 \\
    dust composition$^{a}$  & $\mathrm{Mg_{0.7} Fe_{0.3} SiO_{3}}$ & $95\%$ \\
    (volume fractions) & amorph.carbon &  $5\%$\\
    \hline
    \end{tabular} 
    \tablefoot{$^{(a)}$ Optical constants are from ~\cite{1995A&A...300..503D}. and ~\cite{1996MNRAS.282.1321Z}, BE-sample).}
\end{table}

We model the thermal and chemical evolution of our objects during an accretion burst using the radiation thermochemical code \prodimo \citep{2009A&A...501..383W,2010A&A...510A..18K,2011MNRAS.412..711T}.  {This code provides the temperature structure and local radiation field for the disk density structure and solves the time-dependent chemistry for the provided chemical species. An interstellar background radiation field is also included that mainly shows its impact on the outer parts of the envelope by contributing to the photodesorption in that region of the MYSO.} 
All three models share the main parameters displayed in Table \ref{table:1}. To determine the stellar luminosity and effective temperature of a 10 $M_{\sun}$, 50 kyr old MYSO we used evolution tracks for young stars from \cite{2008ASPC..387..189Y}. 
 {We model the total luminosity of the star by combining a stellar spectrum and a blackbody. The blackbody with a temperature of 15000 K is used to simulate the increase in luminosity by accretion. The stellar spectrum is picked from the PHOENIX catalog \citep{2005ESASP.576..565B} using the properties shown in Table \ref{table:1}.} 
We set the accretion luminosity at $L_{\text{ac}} = 10000 L_\sun$ for the quiescent state and multiplied it by 10 for the burst scenario.  {The burst strength and duration are consistent with \cite{2021A&A...646A.161S} in the case of object G358.93-0.03 and \cite{2022arXiv220311257F} for the cases of DO Tau, EX Lup (2011 and 2008), and V1647 Ori.}
We set the gas temperature equal to the dust temperature. We make this assumption because the regions of interest in the objects are dense enough to assume that the thermal accommodation is efficient enough to keep both temperatures equal.

\subsection{The physical structure}
\label{The physical structure}
For the general physical structure, we use the same approach as in \cite{2017A&A...604A..15R}. For the respective values for the disk and star mass, however, we refer to \cite{2021MNRAS.500.4448M} where a disk-to-star mass ratio of 0.5 is reasonable. 
The models consist of a disk component and, in the case of the two embedded objects, an envelope with an outflow cavity. The density structures of the disk and envelope are calculated separately. The disk structure is then put on top of the envelope wherever the values for the disk density are higher than for the envelope.
{As in \cite{2017A&A...604A..15R} we used the infalling rotating model of \cite{1976ApJ...210..377U}. 
The structure of the envelope is then given by the following equation:}
\begin{equation}
\rho(r,\mu) = \frac{\dot{M}_{\mathrm{if}}}{4\pi}\left(2GM_{*}r^3\right)^{-1/2}\left(\frac{1}{2}+\frac{\mu}{2\mu_0}\right)^{-1/2}\left(\frac{\mu}{2\mu_0}+\frac{2\mu^{2}_{0}R_\mathrm{c}}{r}\right)^{-1},
\end{equation} 
{where} $\dot{M}_{\mathrm{if}}$ is the envelope's mass infall rate, G the gravitational constant, $M_{*}$ the stellar mass,{ $r$ the radius} starting at the star, $R_c$ the centrifugal radius,  $\mu = cos\theta$ is the cosine of the polar angle of a streamline of infalling particles and $\mu_0$ is the respective $\mu$ value for $r \xrightarrow[]{} \infty$. 

We use a fixed parameterized density structure for the disk.  {We refer to  \citet{2009A&A...501..383W,2016A&A...586A.103W} for a detailed description of all the parameters and their implementation in \prodimo.} The gas density structure is an axisymmetric flared (2D) function of the radius and the height of the disk ($r$ and $h$) and it is given by 

\begin{equation}
\rho(r,z) = \frac{\Sigma(r)}{\sqrt{2\pi}h(r)}\exp{\left (  -\frac{z^{2}}{2h(r)^{2}}\right )}\quad   [\text{g cm}^{-3}].
\end{equation} 

{Here,} $\Sigma(r)$ is the radial surface density profile of the disk. We assume a simple power-law distribution with a tapered outer edge

\begin{equation}
\Sigma(r) = \Sigma_{0}\left (\frac{r}{R_{\mathrm{in}}}\right )^{-\epsilon}\exp{\left (-\left (\frac{r}{R_{\mathrm{tap}}}\right )^{2-\gamma}\right )}\quad  [\text{g cm}^{-2}].
\end{equation}

$R_{in}$ is the inner disk radius and $R_{tap}$ is the characteristic radius. The vertical scale height $h(r)$ is given by a radial power law

\begin{equation}
h(r) = H(100\: au)\left (\frac{r}{100\ au}\right )^{\beta}.
\end{equation}

{Here,} $H(100\ au)$ is the disk scale height at $r = 100\ au$ and $\beta$ is the flaring power index. The parameters for the disk density structure are listed in Table \ref{table:1}.  

 {The two models representative of the embedded phase have two different cavity opening angles (30 and 50 degrees) and the model representing the end of the embedded phase consists of a disk without an envelope. The three models represent different stages of a disk. Thus, the model with the smallest cavity angle and the most massive envelope represents the youngest disk, and the model without an envelope is the oldest one in the set. 
 In Table \ref{tab:2} we show the differences between the three models representing different stages of MYSO evolution. It is important to mention that the mass of the envelope is not a parameter but rather the result of how the envelope is constructed in the code.}

\begin{table}[!ht]
\center
\caption{Setup of the envelope of the three different models.}
\label{tab:2}
\begin{tabularx}{0.49\textwidth}{ l|c|c|c } 
 \hline
 \hline
 Quantity & cav30 & cav50 & disk-only \\
 \hline
envelope mass  & 5.2 x10$^{-4}$ & 1.04 x10$^{-4}$  & - \\
infall rate ($\dot{M}_{\mathrm{if}}$)  & $M_{\sun}$/yr & $M_{\sun}$/yr &   \\
 &  &  &   \\
 envelope mass ($M_{\mathrm{env}}$) & 15 $M_{\sun}$ & 3.0 $M_{\sun}$ & - \\
 & & & \\
cavity opening  & $30^{\circ}$ &  $50^{\circ}$ & - \\
angle ($\beta_{\mathrm{cav}}$) & & & \\
 \hline
 \hline   
\end{tabularx}
\end{table}

\subsection{Dust properties}

We assumed a dust-to-gas mass ratio of 0.01 in the disk. We take dust growth into account by using a dust size distribution with a minimum and a maximum dust grain size ($a_{min}=0.05\mu m $ and $a_{max}=1000\mu m$). We used a simple power-law for the dust size distribution $f(a) \propto a^{-a_{pow}}$. We use the canonical value for interstellar grains of $a_{pow} = 3.5$ \citep{1977ApJ...217..425M}. The dust composition consists of a mixture of $95 \%$ amorphous laboratory silicate and $5 \% $ amorphous carbon. All the relevant dust properties are listed in Table \ref{table:1}. In this work, the dust composition and dust size distribution are constant throughout the entire disk. 

\subsection{Chemical model}

The chemical network used in this work is based on \cite{2017A&A...607A..41K}, which in turn is also based on the UMIST 2012 database \citep{2013A&A...550A..36M} for gas-phase chemistry. We additionally use a number of surface chemistry reactions on dust grains described in \cite{2020A&A...635A..16T} and Thi et al. (in prep). The surface chemistry network in ~\cite{2020A&A...635A..16T} is based in \cite{1993MNRAS.261...83H}. 
The total number of species in our network is 238 involving 3500 chemical reactions.
Time-dependent chemistry is used to analyze the evolution of the chemical abundances during and after the burst. The binding energies we use for water and methanol are presented in  Table \ref{tab:3}.  {The difference in binding energies of both molecules is small, hence the freeze-out temperatures and therefore the position of the snowline are very similar.} This is shown more clearly in the Appendix \ref{2D cuts}.
\begin{table}[!ht]
\center
\caption{Binding energies for water and methanol. Both values are taken from \cite{2007MNRAS.374.1006B}}
\label{tab:3}
\center
\begin{tabular}{ l c } 
 \hline
 \hline
 Species &      $E_B [K]$ \\
 \hline
H$_2$O &      4800 \\
 CH$_3$OH &      4930 \\
 \hline   
\end{tabular}
\end{table}

The method we use to establish initial conditions for the abundance is explained in the next section.

\begin{figure}[!h]
    \centering
    \includegraphics[width=1.0\hsize]{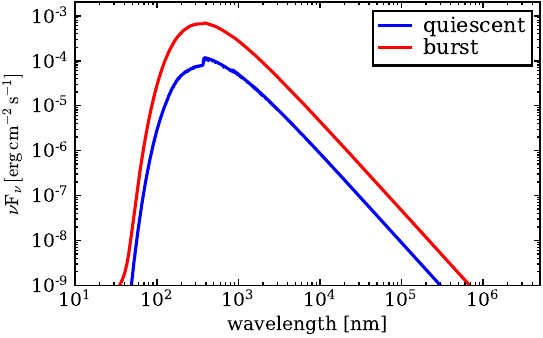}
    \caption{Star spectrum for quiescent conditions (blue) and during the accretion burst (red).}
    \label{fig:starspec}
\end{figure}

 {To be able to consistently model the chemical evolution of a MYSO during and after an accretion burst, we first assume initial element abundances of molecular cloud from \cite{2016A&A...586A.103W}.} {We establish an evolutionary timescale that encloses the formation of the proto-star up to the ZAMS and let the chemistry evolve for that amount of time. We chose a timescale of $5\times10^4$ yrs, consistent with the evolution of a 10 M$_{\sun}$ star {\citep{2019MNRAS.484.2482M}.} We use the resulting chemical abundances as the input for all the positions in the grid for the disk + envelope models and use that output as the pre-burst (quiescent) disk conditions.}

\begin{figure*}[h!]
    \centering
    \includegraphics[width=1.0\hsize]{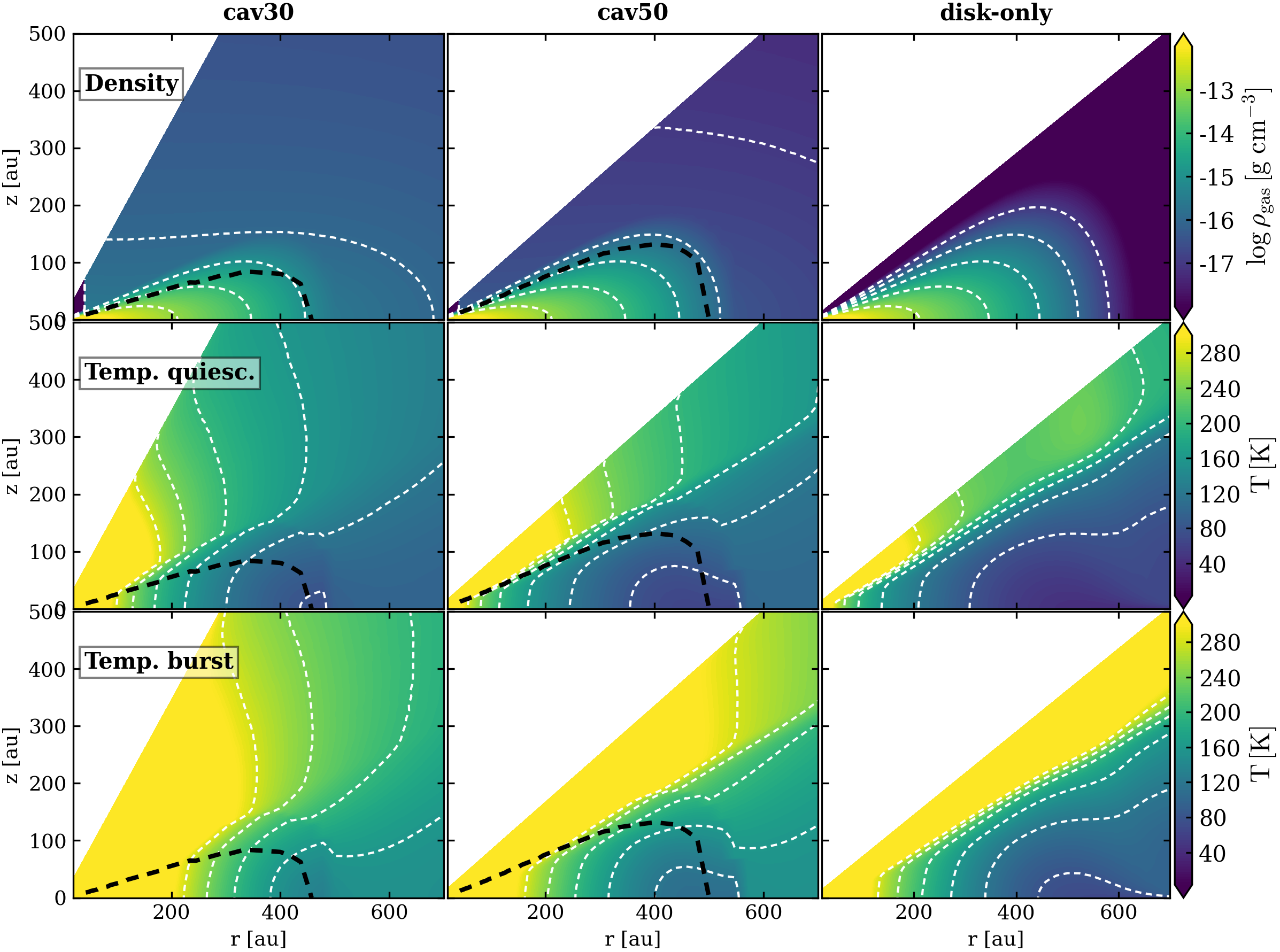}
    \caption{Density (same before, during, and after the burst), and temperature before and during the burst for the three MYSO models. {The temperature structure in the quiescent state (pre-burst) is the same as during the post-burst state.} The model with an opening cavity angle of 30 degrees (l.h.s column), 50 degrees (middle column), and the model without an envelope (r.h.s. column). The white contour lines represent the numbers given in the color bar. The black dashed line marks the transition from the disk to the envelope.} 
    \label{fig:dens_temp_rad}
\end{figure*}

 {As mentioned above, for the quiescent (pre-burst) part, we added the spectrum of the blackbody to the original stellar spectrum. Then we increased the luminosity of the blackbody by a factor of 10 and added it to the original stellar spectrum to simulate the luminosity burst. Fig. \ref{fig:starspec} displays the resulting spectrum of the star used for the quiescent and the post-burst phase (blue) and during the burst (red). The burst lasts for 1 year. The burst duration is consistent within the broad range in burst strength and duration (less than a year to many years) according to the work of  \cite{2021A&A...646A.161S} and \cite{2022arXiv220311257F}. In our particular case, our burst duration is in line with the short burst of G358 ( a 10 M$_\sun$ protostar) studied in \cite{2021A&A...646A.161S}.}  
In the post-burst phase, the accretion luminosity decreases to the quiescent state value $L_{\text{acc}}$ (see Table \ref{table:1}). We focus mainly on the chemical evolution of the disks, therefore we neglect any dynamic changes in the density structure at all times. This is justified because the chemical timescales are shorter than the dynamic timescales of the disk. This means that the density structure is kept fixed at all times.
The steps to simulate the chemical effect of an accretion burst are the following:
\begin{enumerate}
    \item \textit{Pre-burst}. The chemistry of the pre-burst model is set by evolving the chemistry of molecular cloud abundances for $5\times10^4$ years.
    
    \item \textit{Burst}. Local radiation field and temperature structure are calculated with the additional contribution of the burst-accretion luminosity. 

    \item \textit{Burst chemistry}. The chemistry in the disk is evolved during the duration of the burst. For this study, a burst duration of one 1 yr is set.
    
    \item \textit{Post-burst}. The radiation field and temperature structure are recalculated under quiescent conditions.  {The resulting thermal structure is therefore the same as in the pre-burst state.}
    
    \item \textit{Post-burst chemistry}. We take the chemical conditions from the burst and evolve the chemistry further in the post-burst phase for $1\times 10^2$ years under quiescent conditions.  
\end{enumerate}

\section{Results}
\label{Results}
 {In this section, we present the results of our three different MYSO models.
We show the density and temperature distribution as well as the vertically averaged abundances of CH$_3$OH as a function of the radius. Additionally, we show the interplay of ice and gaseous methanol. We show these results prior to, during, and after the burst. We also show the same type of results for H$_2$O in Appendix \ref{h2o_evo}.}

\subsection{Physical structure}\label{Phys str}
 {The top row of Fig. \ref{fig:dens_temp_rad} displays the density distribution of each model. For the disk-only model (right panel), the absence of an envelope is evident as the density strongly drops in regions outside of the disk. The difference in the opening cavity angle in models cav30 and cav50 (left and middle panel) is also visible. The dashed black contour line represents the transition between disk and envelope.} 
The temperature structure of each disk model is also shown in Fig. \ref{fig:dens_temp_rad} for the quiescent state (the middle row) and burst (bottom row). The model with the most massive envelope (cav30) has the highest disk temperature. However, the envelope presents another scenario. While for radii inside r = 200 au the most massive envelope does show the highest temperature, for radii beyond r = 200 au the opposite happens. A more massive envelope will have a steeper radial temperature gradient and therefore lower temperature farther out than a less massive one.
The different temperature trends are the result of two different processes.
For the disk, radiation scattering is the process that leads to higher temperatures for more massive envelopes. If the envelope is more dense and covers a broader area above the disk, it will be more likely that photons coming from the star will be scattered to the disk, providing extra heating.
At distances greater than 200 au, radiation shielding is the process that leads to a lower temperature for more massive envelopes.
This is valid for the two conditions that are presented here: the quiescent (pre-burst) state conditions and the burst conditions.

\begin{figure*}[h!]
    \centering
    \begin{subfigure}[t]{0.33\textwidth}
    \includegraphics[width=\textwidth]{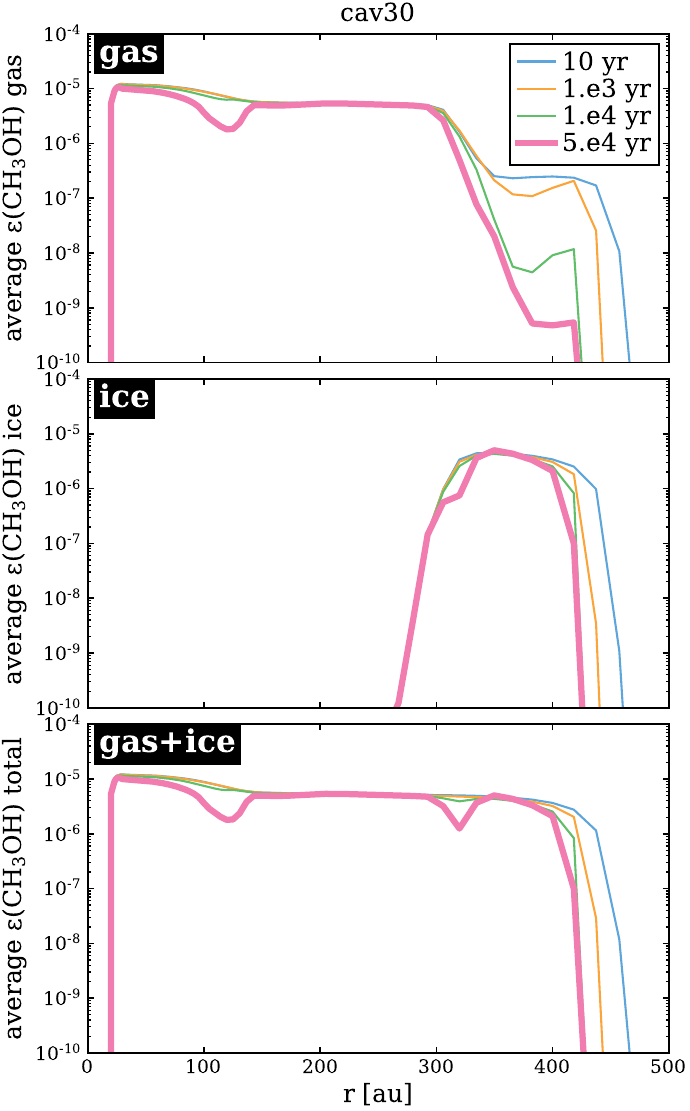}
    \end{subfigure}
    \begin{subfigure}[t]{0.33\textwidth}
    \includegraphics[width=\textwidth]{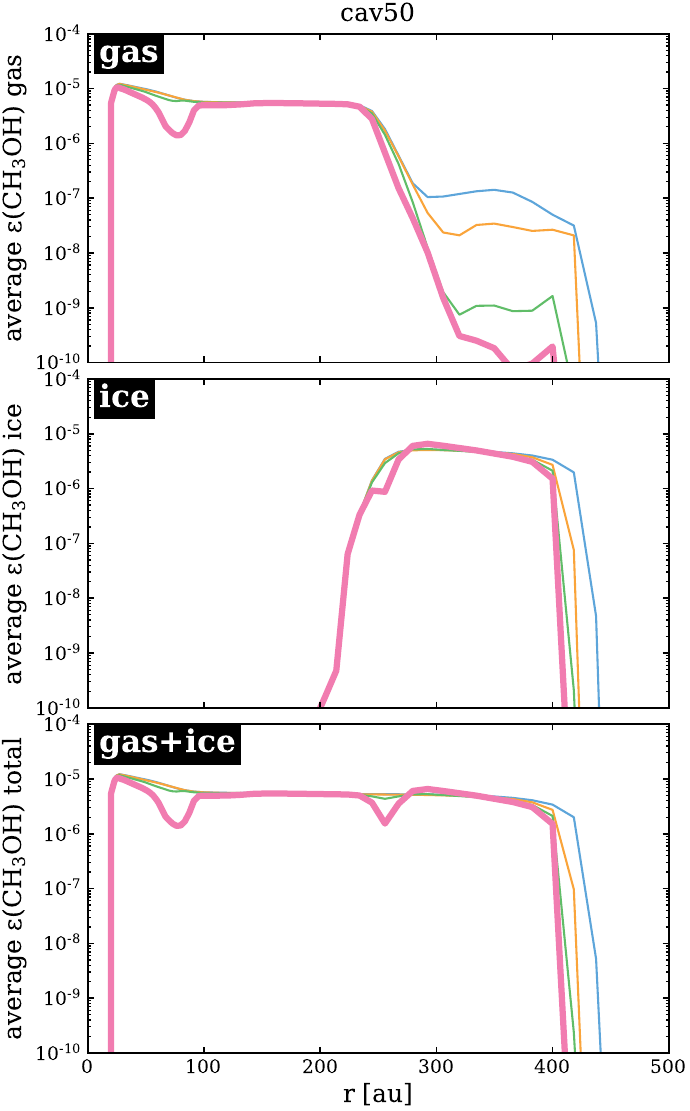}
    \end{subfigure}
    \begin{subfigure}[t]{0.33\textwidth}
    \includegraphics[width=\textwidth]{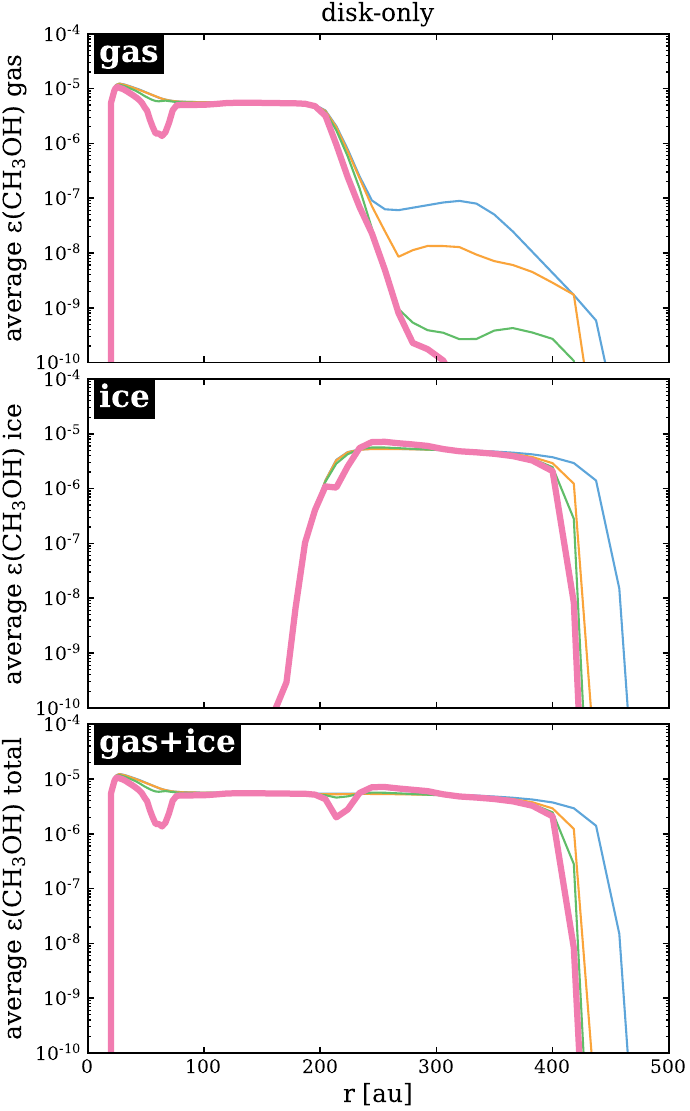}
    \end{subfigure}
    \caption{Snapshots of the evolution of the vertically averaged methanol abundance before the burst scenario. The left, middle, and right columns represent models cav30, cav50, and disk-only respectively. The top, middle, and bottom rows show the methanol gas, ice, and total abundance respectively.}
    \label{fig:init_cond}
\end{figure*}

\begin{figure*}[!ht]
    \centering
    \includegraphics[width=1.0\hsize]{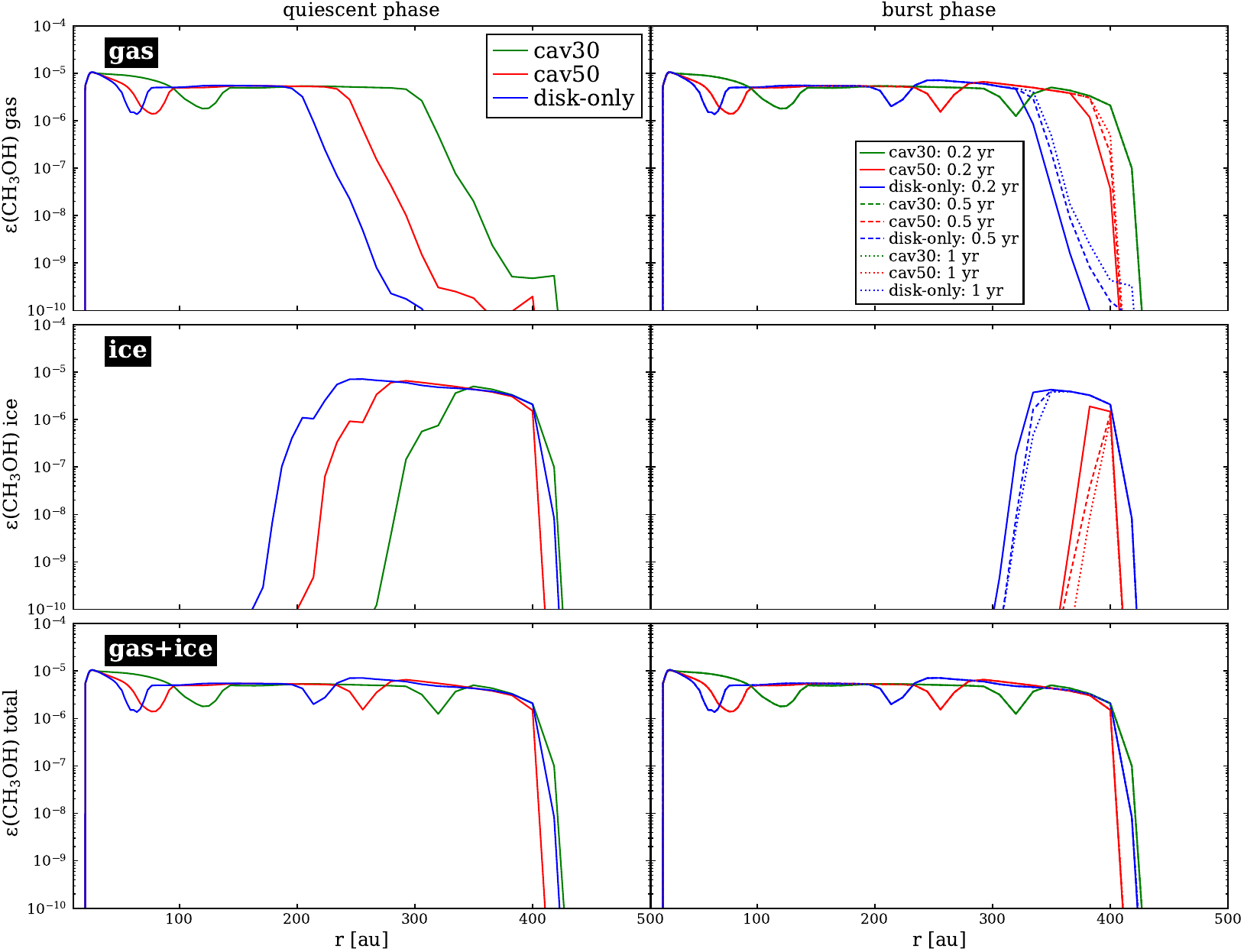}
    \caption{Averaged abundances of methanol for models cav30 (green), cav50 (red), and disk-only (blue). The left and right columns show the pre-burst phase and the burst phase respectively. Three different time steps during the burst are displayed. The top, middle, and bottom rows show the averaged gas, ice, and total abundances respectively.} 
    \label{fig:all_all}
\end{figure*}

 \subsection{Initial chemical structure}\label{Initial chemical structure}
  {We show the vertically averaged abundances as the main tool to display the chemical evolution in the disk and envelope. This value is a result of dividing the vertical column number density $N_{\mathrm{sp}}$ of the molecule of interest by the hydrogen vertical column number density.
To display the ice-to-gas ratio for the complete object we integrated the vertical column number density of the gas and ice phase separately over the radius $r$. 
We used a simple integration from the inner radius $R_{\mathrm{in}}$ to the outer radius $R_{\mathrm{out}}$ of the disk for this step. Since the cells in the grid have a finite size, the integral presented below is actually a sum over all grid cells. 
\begin{equation}\label{integ}
n_{\mathrm{tot}}=2\pi \int_{R_{\mathrm{in}}}^{R_{\mathrm{out}}} N_{\mathrm{sp}} \;r \,dr  ,    
\end{equation}
Once we have $n_{\mathrm{tot}}$, we calculate the ice-to-gas ratio with Eq. \ref{ratio}.
\begin{equation}\label{ratio}
\mathcal{R}= n_{\mathrm{tot(ice)}}/n_{\mathrm{tot(gas)}}.
\end{equation}}

{Fig. \ref{fig:init_cond} shows the methanol (gas, ice and gas+ice) evolution during the first $5\times10^4$ yrs before the burst for each model. 
 {The evolution of the methanol abundances is similar for all models in the sense that as the chemistry in the disk evolves only the outer parts of the disk show a decrease in gas and ice abundances.} However, the radius where the gas abundances start to decrease is different for each model. This radius decreases with decreasing envelope mass. It is around 300, 250, and 200 au for model cav30, cav50, and the disk-only model respectively.  {This is a result of the disk being colder as the envelope surrounding it becomes less massive.
Such a trend is not as clear for the ice abundance or for the total abundance (gas + ice). Although there is a change with time beyond 400 au, in both cases, ice and total abundances, there is a clear cut-off at a radius between 400 and 500 au for all three models. This cut-off is found at decreasing radius as the chemistry in the disk evolves. This decrease of the total abundances is mainly the result of dissociation processes in the outer parts of the envelope caused by the external radiation field.} 
All the models share a drop in the gas methanol abundances inside a radius of 150 au for the last time step. The shown profiles suggest that the drop has a correlation with the envelope. For a model with a less massive envelope, the methanol drop is less strong. It is also less extended and the radius where the drop starts is smaller for a less massive envelope. However, a deeper look into this drop also present in other studies using the same code leads to the conclusion that it is a result of a numerical error and not a physical effect.   {It should be also mentioned, that this feature is not relevant to this study as we are mainly interested in the evolution of snowlines in outer regions of the object.}

\subsection{{Evolution of methanol}}\label{chem_evo}

\begin{figure*}[!ht]
    \centering
    \includegraphics[width=1.0\hsize]{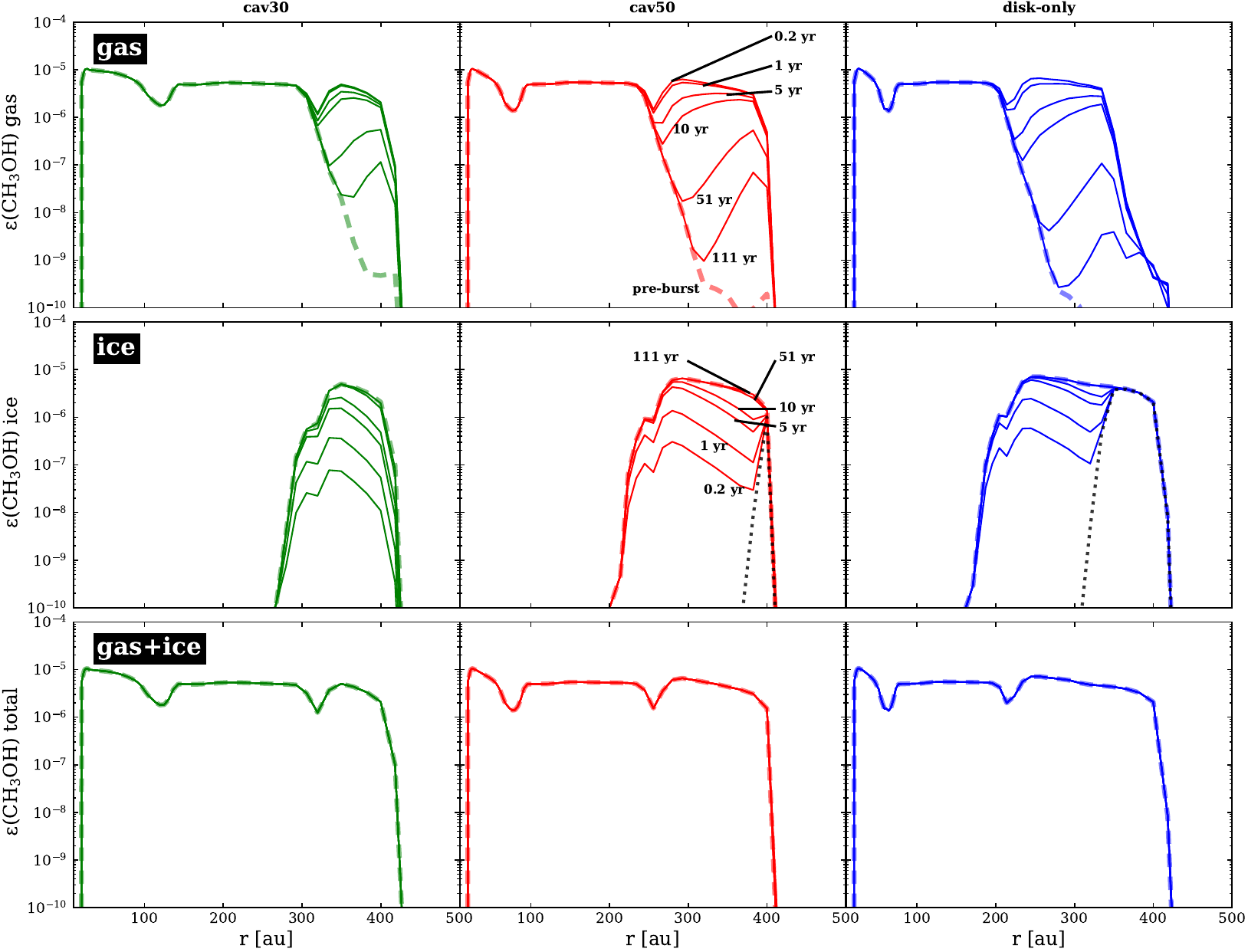}
    \caption{Averaged abundances of methanol for models cav30 (green, left column), cav50 (red, middle column), and disk-only (blue, right column) during the post-burst phase. Six different time steps during the post-burst are displayed. The top, middle, and bottom rows show the averaged gas, ice, and total abundances respectively.  {The black dotted line in the center and the center-right panels show the averaged ice abundance at the end of the burst.}} 
    \label{fig:all_pb}
\end{figure*}

In this section, we present the results concerning the impact of the accretion burst on CH$_3$OH. 
We follow the changes in the vertically averaged abundances as a function of the radius of CH$_3$OH during these processes (Figs. \ref{fig:all_all}, and \ref{fig:all_pb}). As mentioned above, we use the vertically averaged abundances because they offer a simple and clear view of the time evolution of the chemical species. We note that using the averaged abundances could lead to miss changes that occur throughout the vertical extent of the disk. In order to avoid this we also show 2D plots displaying some snapshots of the abundances as a function of radius and height in Appendix \ref{2D cuts}. 
We also follow the ice-to-gas ratio of methanol (Fig. \ref{fig:t_ice_gas_ratio}).
Additionally, we show the evolution of the snow region graphically in a vertical disk cross-section in Fig. \ref{fig:ch3oh_ice_gas}. 
 {The snow region is the region of the disk where a particular species is mostly found in a frozen state. As the disk has a radial and a vertical temperature gradient, we refer to the snow surface as the surface that separates the regions where the ice abundances of a chemical species are larger than the gas abundances from the region where the gas abundances are dominant. Consequently, the snowline corresponds to the position of the snow region at the midplane of the disk.}

Figs. \ref{fig:all_all} and \ref{fig:all_pb} display the evolution of the average abundances of methanol as a function of radius for all models.  {Fig. \ref{fig:all_all} corresponds to the pre-burst (left panels) and the burst (right panels) conditions, while the chemical evolution after the burst is shown in Fig. \ref{fig:all_pb}.}
There are some trends that are shared by the three models. For instance, the total average abundance (gas + ice) in the disk is not affected by the burst (bottom row).    
The top and middle rows of the same figure show instead a clear response of the ice and gas phase to the burst. This indicates that the occurrence of a burst mainly enhances the thermal desorption of frozen species from dust grains into the gas phase.
During the burst, methanol gets strongly desorbed and, consequently, the ice abundances drop by many orders of magnitude and the gas abundances increase also by a couple of orders of magnitude (see Figs \ref{fig:all_all} and \ref{fig:all_pb}). As the post-burst phase starts, the gas and ice abundances return to their initial, pre-burst, values.  {Although the top panels in Fig. \ref{fig:all_pb} show that 111 years after the burst the outer parts of the objects still have higher values for the gaseous methanol than in the pre-burst phase, the difference with the values during the burst is around two or more orders of magnitude. Hence, it is safe to consider that the pre-burst conditions are reached after approximately 100 years. }
For this reason, we decided to leave the averaged abundances of later time steps out of these results for all three models. Another reason for the exclusion of later time steps is that for post-burst times of e.g. $1\times 10^3 - 5\times 10^5$ years we would expect subsequent burst to occur and also a dramatic change of both mass and physical conditions of disk and envelope.

However, differences between the models are also present. For model cav30 the burst only affects the region beyond 300 au, which represents the outer part of the disk. This could be explained by the fact that inside 300 au most molecules are already in the gas phase. Therefore there is not much ice inside 300 au to be desorbed when the burst sets in. For radii outside 300 au, the ice reservoir of methanol gets strongly depleted. This leads to the strongest drop in the ice-to-gas ratio in Fig. \ref{fig:t_ice_gas_ratio} (top panel) and for the vanishing of the methanol snow region in Fig. \ref{fig:ch3oh_ice_gas} during the burst for model cav30

Regarding model cav50, there are two visible differences from model cav30. First, the impact of the accretion burst on the gas phase of methanol sets in for a smaller radius than model cav30. In this case, the effect is already noticeable for radii beyond 250 au. This is a consequence of the lower temperatures in the disk embedded in a less massive envelope. Molecules in the ice phase are able to survive in regions closer to the star until the burst takes place and desorbs the molecules from dust grains.
Second, the ice abundances do not experience the same depletion during the burst in the outer parts of the disk. This is so because of the lower temperatures present in the cav50 model. Even during the burst, the temperature does not increase enough to provide the same amount of ice depletion as in more heavily embedded disks. This is shown in the middle row of Fig. \ref{fig:all_all}. The blue lines that represent the abundance during the burst show a strong ice depletion except for a radius between 360 and 410 au. This is also visible in Fig. \ref{fig:ch3oh_ice_gas}, where the frozen methanol abundance region is displayed. 
Figs. \ref{fig:t_ice_gas_ratio} and \ref{fig:ch3oh_ice_gas} indeed show a weaker degree of ice depletion for the cav50 model and a remaining snow region around a radius of 400 au.

The third model ("disk-only") shows that the radius where the impact of the burst becomes noticeable is around 200 au. The effect of the burst on the ice abundances is almost not present for radii between 300 and 420 au. In this case, the effect of the burst is limited in this region because, even during the burst, the temperatures needed for thermal desorption are not reached in that part of the disk.
The drop of the ice-to-gas ratio during the burst is consequently also the weakest for this model (Fig. \ref{fig:t_ice_gas_ratio}) and the remaining snow region is also the most extended of all models.
 
This is consistent with the differences in temperature, as an object with higher temperatures will favor thermal desorption.   
As the burst starts, the ice-to-gas ratio $\cal R$ shown in Fig. \ref{fig:t_ice_gas_ratio} (top panel) for the "disk-only" model drops by at least one order of magnitude and continues to decrease in a much weaker manner. The ratio for model cav50 also drops by approximately two orders of magnitude and continues decreasing more rapidly than the "disk-only" model till the end of the burst. For model cav30 the burst leads to a dramatic drop of the ratio of approx eight orders of magnitude, which remains mostly constant during the burst. As mentioned above, after the burst, the three models go back to the initial ice-to-gas ratios at the same rate and reach pre-burst ratios at $\approx 100$ years after the end of the burst. 
 
\begin{figure}[!h]
  \centering
  \begin{subfigure}[t]{0.98\hsize}
\includegraphics[width=\hsize]{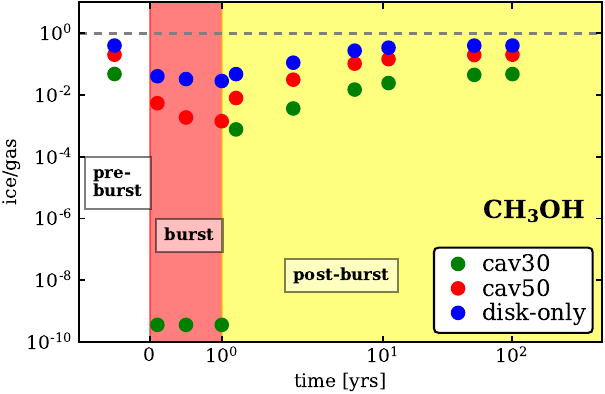}
    \end{subfigure}
    \begin{subfigure}[t]{0.98\hsize}
\includegraphics[width=\hsize]{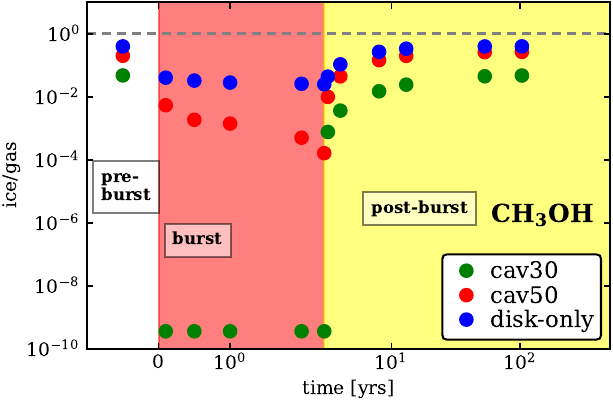}
    \end{subfigure}
\begin{subfigure}[t]{0.98\hsize}
\includegraphics[width=\hsize]{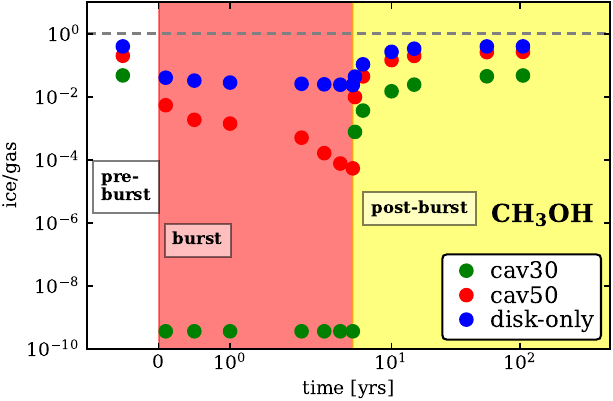}
    \end{subfigure}
\caption{ Ice-to-gas ratio evolution over time. The white, pink, and yellow regions represent the pre-burst, burst, and post-burst phases respectively. The green (model cav30), red (model cav50), and blue (model disk-only) dots show the ice-to-gas ratio as a function of time. The top, middle, and bottom panels show the ice-to-gas ratio of methanol for the setups discussed so far for a burst duration of 1, 3, and 5 years, respectively.}
\label{fig:t_ice_gas_ratio}
\end{figure}

In Fig. \ref{fig:ch3oh_ice_gas} the different extensions of the respective snow regions also support the idea that the more massive an envelope is around a disk, the stronger will be the impact of the burst on the methanol ice abundance.

For all three models, the timescale to replenish the snow region has also a value of $\approx$ 100 years. This timescale is important because it defines the time range in which chemical signatures caused by the burst can be detected \cite[]{2017A&A...604A..15R}. The adsorption timescale can be written in the following form according to \cite{2017A&A...604A..15R}:
\begin{equation}
    t_{\text{i,ads}}= 2.9\times 10^{-12}M_{\text{i}}^{1/2}\alpha^{-1} \frac{\rho_{\text{dp}}}{\delta}T_{\text{gas}}^{-1/2}\frac{\langle a^3 \rangle}{\langle a^2 \rangle} \rho_{\text{gas}}^{-1} \ \ [\text{yr}]
\end{equation}
with $M_\text{i}$ being the molecular weight of species $\text{i}$, $\alpha$ the sticking efficiency, $\rho_{\text{dp}}$ the material density of a dust particle, $\delta$ the dust-to-gas ratio, $T_{\text{gas}}$ the gas temperature, $\langle a^3 \rangle$ and $\langle a^2 \rangle$ the third and second moment of the dust size distribution respectively, and $\rho_{\text{gas}}$ the gas density.

Thus the freeze-out timescale is, in part, gas density-dependent. This means that higher density will result in shorter timescales. Therefore the freeze-out timescale will decrease as the radius decreases. The resulting effect will be the inside-out freeze-out shown in Fig. \ref{fig:ch3oh_ice_gas}.  {The different temperatures lead to a different extension of the snow region for each model and the negative density gradient is responsible for the inside-out freeze-out of the species.} All models share the same disk structure, therefore the density in the disk is almost exactly the same in all models. The similar density for all the models in the region where the ice species are being sublimated and frozen out again is responsible for the similar freeze-out timescales.

\begin{figure}[!h]
\centering
    \includegraphics[width=1.0\hsize]{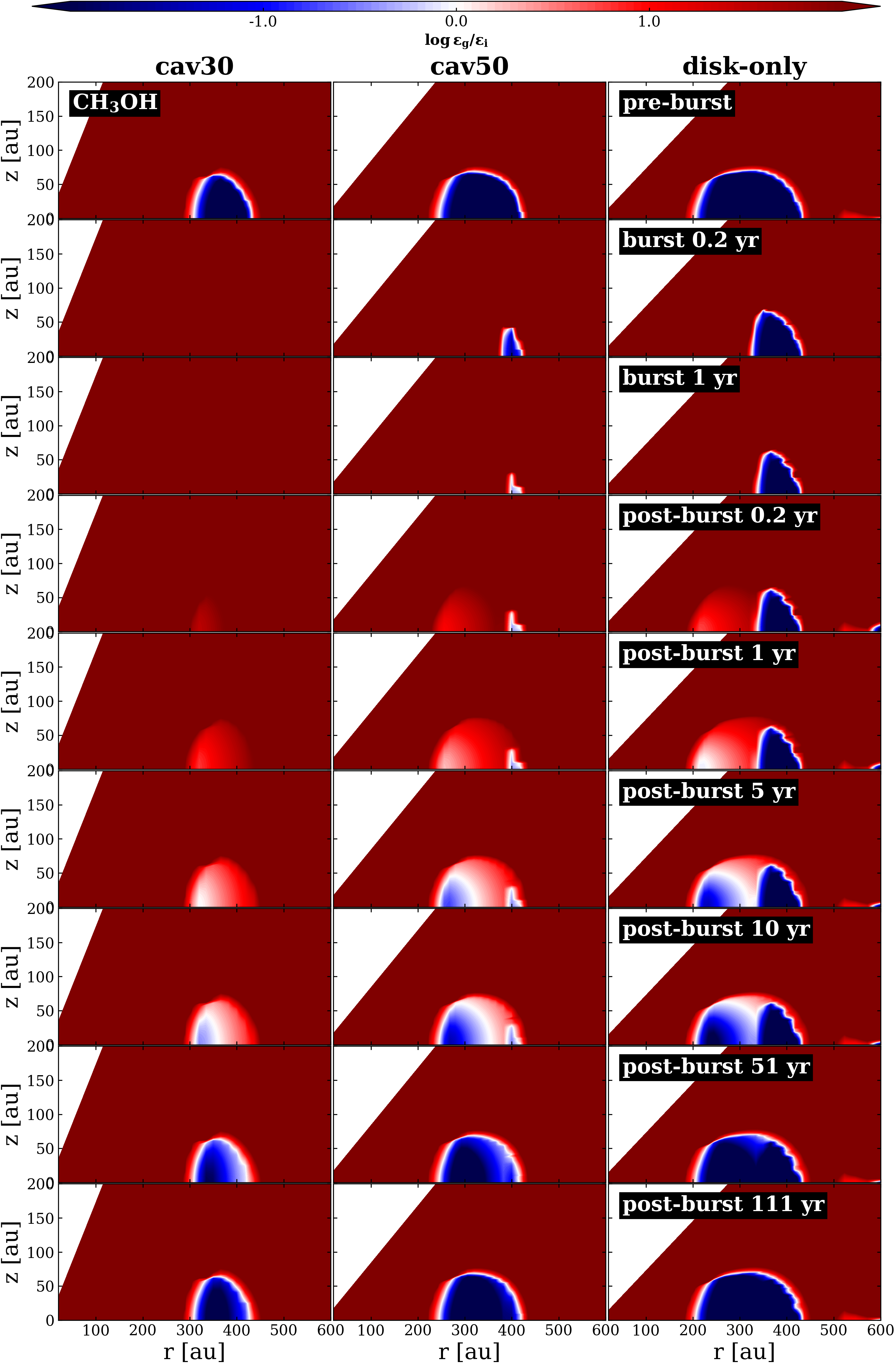}
    \caption{Ice to gas methanol ratio as a function of the radius and height of the disk. The blue and red regions correspond to the regions dominated by ice and gas respectively. The white regions inside the disk represent the snow surface that encloses the snow region of the disk. The left, middle, and right columns represent cav30, cav50, and disk-only models respectively. Each row stands for a snapshot in their evolution before, during, and after the burst.}
    \label{fig:ch3oh_ice_gas}
\end{figure}

  {Although the objects in the studies mentioned below are low-mass stars, they provide a couple of freeze-out timescales that enable us to gain a sense of the ice replenishment process.}
 For instance, the reformation timescale of H$_2$O ice is leveraged at 1,000 years and at 10,000 years for CO for a sample of Class 0/I sources  {(\citealt{2019ApJ...884..149H} based upon the work of \citealt{2012ApJ...754L..18V}).} In \cite{2018ApJ...866...46M}, H$_2$CO, and NH$_2$OH have a freeze-out timescale between 10 and 1,000 years. We note that as the freeze-out timescale mainly depends on the density and our models have higher densities than low-mass stars, the expected freeze-out timescales are expected to be shorter.

\begin{figure*}[!ht]
    \centering
    \includegraphics[width=1.0\hsize]{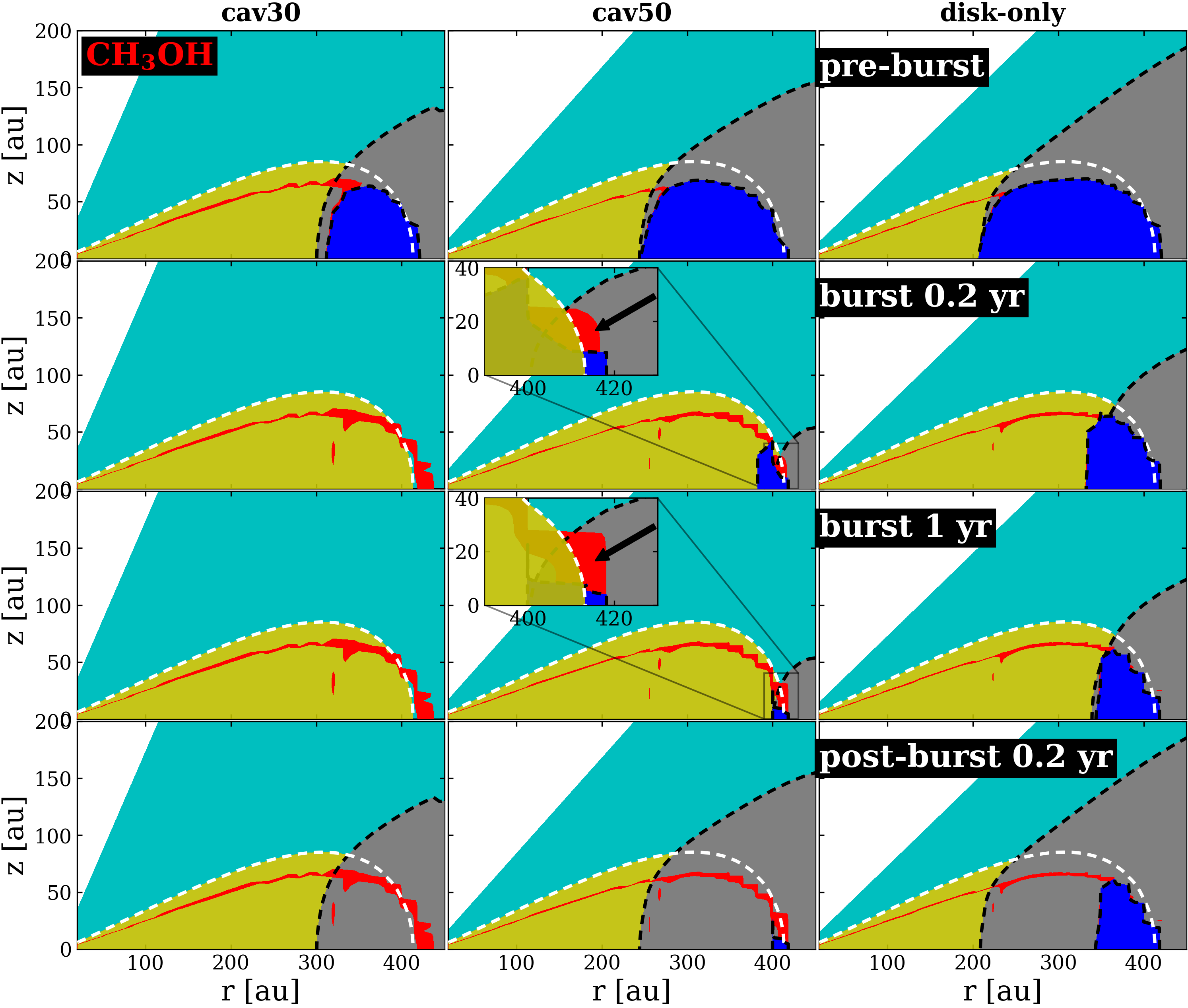}
    \caption{Different zones where the conditions (dust temperature and H nuclei density) for methanol masers are fulfilled or are not fulfilled. The zones shown in different colors are as follows: grey region: proper temperature, enclosed by the black dashed lines. {In regions with lower temperatures ethanol freezes onto dust grains. Higher gas temperatures increase the probability of collisional de-excitation of the methanol molecules.}  
    The critical value for gas density is $10^9 \text{cm}^{-3}$(white dashed line). Below the white dashed line the gas density is too high for masers. {Dark blue region: frozen methanol}. The red region shows the regions in the disk where the methanol gas abundance is optimal for maser occurrence. Thus, the overlap of the grey region and the red region (above the white dashed line) is the region where methanol masers could occur.  Two zoomed-in regions are added in the middle column to indicate where methanol masers are possible i.e. the region where temperature, density, and gas methanol abundance favor masers.}
    \label{fig:maser_cond}
\end{figure*}

\subsection{Ocurrence of methanol masers}\label{masers}

Additionally, to the results presented above, we tested our models regarding the occurrence of methanol masers. Flares of methanol masers are driven by luminosity bursts in MYSOs. Therefore, the occurrence of masers can be used to track luminosity bursts. The fact that methanol masers can be associated with luminosity bursts is mentioned in \cite{2021A&A...646A.161S} where it is stated that for two MYSOs (S255IR-NIRS3 and NGC6334I-MM1) a luminosity increase in the infrared and (sub)mm regime was seen. This increase is thought to be due to higher accretion rates \cite[]{2017MNRAS.464L..90M}. Moreover, the outbursts detected were accompanied by flares of methanol masers. This confirmed that the maser flares were a result of radiative pumping produced by the enhanced luminosity during the accretion burst.

In Fig. \ref{fig:maser_cond} we located the regions in the disk where the conditions that favor methanol masers are present for time steps before, during, and after the burst in our simulations.  {The proper temperature and hydrogen nuclei density conditions are in part taken from \cite{2021A&A...646A.161S}.} Fig. \ref{fig:maser_cond} shows the proper temperature regime between the minimum temperature for thermal desorption of methanol and 120 K (optimal temperature for methanol desorption) enclosed by the black dashed lines. The proper gas density regime is the region above the white dashed line. {The highest temperature contour (120 K) is relevant because for higher gas temperatures the probability of methanol masers decreases. In a region with higher gas temperatures, the methanol molecules gain kinetic energy and are therefore more susceptible to collisional de-excitation.}

The value for the maximum gas density in \cite{2021A&A...646A.161S} is of $10^8 \text{cm}^{-3}$. However, other studies \citep{2005MNRAS.360..533C} have suggested that the value could be increased by up to one order of magnitude. Although \cite{2021A&A...646A.161S} state that exceeding the density of $10^8 \text{cm}^{-3}$ would lead to a rapid drop in maser brightness due to collisional de-excitation,
we decided to adopt a maximum gas density value of $10^9 \text{cm}^{-3}$ because using the maximum value of $10^8 \text{cm}^{-3}$ would eliminate the possibility of masers in all our models. Moreover, we note that we use a fixed-density structure in our model. This means that we neglect any changes in the density distribution of our models that would take place in the case of an accretion burst (see, e.g. \citealt{2020A&A...643A..13V}). The changes in the density distribution would alter the regions in the object where the proper gas density for methanol masers is fulfilled. Thus, proper density for methanol masers could be found for lower disk heights and lower radii.
According to the temperature and gas density, the grey region above the white dashed line in Fig. \ref{fig:maser_cond} is the only region in the disk where the optimum temperature and density conditions are met. 
Additionally, the regions of the disk where gas-phase methanol has optimum abundances for maser excitation (between $10^{-8}$ and $10^{-6}$) are displayed in red. Thus, the locations above the white dashed line where the red and the grey regions overlap are the regions where methanol masers are expected to occur. According to our results, only the cav50 model will exhibit methanol masers during the accretion burst. These possible masers are located in the outer parts of the disk beyond 400 au. Therefore, according to these results, the occurrence of methanol masers depends mostly on the optimum combination of the envelope and burst strength to produce high enough temperatures to desorb methanol in regions of the disk where the gas density is below the maximum gas density.

 \section{Discussion}\label{Discussion}

 As presented in the results, the time that the chemistry in our three models requires to recover from the burst (i.e. reach pre-burst conditions) is at the most 100 years. Furthermore, most of the change occurs in the first 10 years after the burst. This means that with massive stars it becomes more feasible to observe this process at different stages. A possible scenario would be to first detect a burst. This would trigger observations (e.g. with The Atacama Large Millimeter Array, ALMA) once a year. The data gathered from this could help us gain insight into the chemistry (i.e. adsorption processes) in general and also potentially into the dust properties in the object.
 {In contrast, typical bursts in low-mass stars are usually much longer and therefore less likely to be observed in the different stages. Shorter bursts in low-mass stars tend to be weaker and only affect the disk at small radii, which makes them more difficult to observe.
 However, the fact that massive stars tend to be farther away poses a challenge to estimating the resolution needed to resolve certain features in the object (e.g. the location of the snowline). Models like the ones presented in this study would then be crucial to interpreting the data obtained from the observations and/or providing predictions for timescales. With \prodimo, we are able to make predictions for synthetic observables as well. This will be addressed in future papers.}
 
  As already mentioned, these models assume certain simplifications. Therefore, it is important to keep in mind that if the models would include other physical effects the structure of the disk would be altered and as a consequence, the chemistry would be influenced as well. For example, the models used in this work are passive disks. This means that the only source of energy in the object is the radiation field produced by the star and the cosmic rays. If the models would take viscous heating due to turbulences in the disk into account, the density and temperature structure of the object would be different. In some parts of the object, the density would sink and the temperature would rise. This would in turn also change the regions where certain chemical processes take place. As a consequence, the chemical structure would also change. In relation to the methanol masers, this would mean that the methanol maser region could be located at lower altitudes in the object and at smaller radii. {We also note that the density distribution for strongly gravitationally unstable disks (GI-unstable disks) around massive protostars is typically non-axisymmetric. This has been inferred, for example by \cite{2022MNRAS.517.4795M} and \cite{2020ApJ...904..181L}using modelling and observations. Non-axisymmetry can introduce very different density structures in comparison with the models presented in this paper (e.g. local depressions between spiral arms, etc...).}

\subsection{Ice-to-gas ratio for longer bursts}\label{Ice-to-gas ratio} 
{We performed the same simulations presented here with longer-lasting bursts to gauge the impact of the burst's duration. The middle and bottom panels of Fig. \ref{fig:t_ice_gas_ratio} display the ice-to-gas ratio of methanol under the influence of an accretion burst with a duration of 3 (middle panel), and 5 years (bottom panel).
These results show that after approximately 3 years of burst conditions, the ice-to-gas ratio does not decrease significantly for the disk-only and the cav30 model. Model cav50, however, shows a clear decrease until the end of the burst. In other words, the main impact of the burst on the desorption processes takes place in the first three years of the burst for the most heavily embedded model and for the model without an envelope but not for the model in between. 
This suggests that the remaining snow region of model cav50 seen in Fig. \ref{fig:ch3oh_ice_gas} is further depleted for the burst strength used in this study if the burst lasts longer than 1 year. Model cav30 has already depleted most of the ice during the one year of burst conditions and the disk-only model will keep the remaining snow region unaffected even if the burst last 4 years longer. Another finding is that the freeze-out timescale in the post-burst phase seems to be independent of the burst duration for all models. In order to confirm this behavior, studies with time-dependent radiative transfer should be performed. The chemical evolution is traced up until $\approx$1000 years after the burst to confirm that the ice-to-gas ratio does not change after returning to pre-burst values.

 We also show some of the same results for H$_2$O in Appendix \ref{h2o_evo} and offer a short interpretation of the results. Overall, H$_2$O behaves very similar to CH$_3$OH. However, some differences arise due to the photodissociation of H$_2$O, as will be discussed in more detail in the Appendix.}

\subsection{Comparison with other studies}\label{Comparison with other studies}
 {We compare our results with the ones provided in \cite{2021A&A...646A.161S} for the source NGC6334I-MM1. In this work a stellar mass of 12 $\pm$ 3M$_{\sun}$ is derived. A methanol maser ring during the outburst observed in \citep{2020NatAs...4..506B} is located at a radius of approximately 900 au, which in the case of this source is still inside the outer radius of the disk. As mentioned above, in the case of our models we find for model cav50 a region for methanol maser at an approximate radius of 415 au. The discrepancy between the model and observation represents a motivation to vary a couple of basic parameters that would mimic physical effects that determine the state of the modeled objects.
In order to compare models with other observed methanol masers, models with similar physical characteristics should also be produced. For example, methanol maser emission has been observed at tens of au from the source in (\citealt{2017A&A...600L...8M}). In order to reproduce such maser emission, our models should consider using a less dense disk and/or environment when defining the physical conditions of the model among other less relevant factors. This could enable us to perform a qualitative comparison on the radial motion of the maser emitting region seen by observations.}

\section{Summary and conclusions}
\label{Summary and conclusions}
We used the thermochemical code \prodimo to simulate the impact of an accretion burst on the circumstellar environment of MYSOs. For this work, we tested three models of a 10 M$_{\sun}$ MYSO with different types of envelopes. The structure of two models represents less evolved ($\approx$10$^4$ yr) sources, which stands for the time where the star and disk are still embedded in an envelope and a third model disk-only model represents the phase where the envelope is already depleted (5$\times 10^4-10^5$ yr) but the disk is still present. The two embedded models differ in the opening cavity angle of the envelope and consequently, in the amount of mass that the respective envelope contains. 
Our main findings are:

\begin{itemize}
    \item The degree to which the disk is embedded in an envelope has an impact on the temperature of the disk and envelope. The temperature in the disk will increase if the disk is more heavily embedded. This is a consequence of the photons coming from the star being scattered to the disk and heating it up. The probability of this happening is higher for envelopes that cover more space around the disk (i.e. a smaller opening cavity perpendicular to the disk). For the temperature in the envelope, the situation is different. For small radii (< 100 au) the envelopes share similar temperatures. However, for the most distant parts of the envelope, the amount of mass in the envelope is able to absorb the stellar radiation and shield the rest of the envelope. This leads to lower temperature values in the outer envelope for more heavily embedded objects. The general impact on the chemical species is that a less embedded object will have a larger amount of species frozen onto dust grains.
    
    \item {Because of the different disk temperatures that lead to a different amount of ice reservoir, the scenarios during the burst also differ. The more embedded an object is, the less extended the affected region by the burst will be. The most heavily embedded model presents the smallest radius interval where the methanol and water abundances will show a signature of the burst (300 to 450 au). As the object becomes less embedded, the affected radius interval will extend inwards into the disk. The disk-only model shows that the impact of the burst is already noticeable at 180 au. The reason for this is the different amounts of ice present in the disk, which is larger for less embedded sources. Therefore the impact of the burst will be noticeable in regions in the disk where ice is present and can be desorbed by the burst.
    Additionally, desorption and adsorption are the most dominant processes that drive the impact of the accretion burst. There is no clear impact of the burst on the total (gas + ice) abundances. This is also shown in the 2D abundance distribution in Appendix \ref{2D cuts}. Therefore no additional destruction or formation reactions are strong enough to be registered.} 
    
    \item {During the burst, the snow region of embedded objects is heavily depleted or completely vanishes. In the case of the disk-only model, there is a portion of the snow region that remains mostly unaffected even during the burst. The most heavily embedded model shows the strongest ice depletion during the burst (more than 8 orders of magnitude), followed by a less embedded model ($\approx$ 2 orders of magnitude) and the disk-only model ($\approx$ 1 order of magnitude).}
    
    \item {All three models share an ice replenishment timescale of $\approx$ 100 years for H$_2$O and CH$_3$OH. This suggests that if the disk is the same and only the envelope is different, the time for a disk to return to its pre-burst state is independent of the envelope. The snow region is replenished from the inside out in all three cases due to the density gradient. In the disk-only model, the surviving portion of the snow region merges with a growing snow region reemerging from denser parts of the disk. The embedded models do not exhibit a surviving snow region at the end of the burst. Instead, they replenish the snow region with the process of an inside-out freeze-out until the initial snow region extension is reached. The only difference between the two latter models is the fact that with a more massive envelope present, the replenishment starts at a larger radius.}  
\end{itemize}

This study opens up the possibility of detecting new and past accretion bursts in MYSOs using gas and ice abundances as well as snowline positions. Together with the new observing possibilities available with the JWST to measure these quantities, we could gain more insight into the pre-main sequence evolutionary phase of MYSOs.
For the moment this comparison would be limited to past accretion bursts within $\approx$ 100 years after the burst and to the gas and ice features that can be detected with the JWST. The chemical modeling could also be expanded to study the features of chemical species beyond CH$_3$OH and H$_2$O.

\begin{acknowledgements}
 We thank the anonymous referee whose comments helped to improve the quality of the manuscript.
 The computational results presented have been achieved using the Vienna Scientific Cluster (VSC). This work was supported by the Austrian Science Fund (FWF) under research grant P31635-N27. Ch. Rab is grateful for support from the Max Planck Society and acknowledges funding by the Deutsche Forschungsgemeinschaft (DFG, German Research Foundation) - 325594231. MG acknowledges support from the Max-Planck-Institute for Astronomy (Heidelberg) during his sabbatical visit. A.C.G. has been supported by PRIN-INAF MAIN-STREAM 2017 “Protoplanetary disks seen through the eyes of new-generation instruments” and PRIN-INAF 2019 “Spectroscopically tracing the disk dispersal evolution (STRADE)”. A.M.S. acknowledges support from the Ministry of Science and Higher Education of the Russian Federation by an agreement FEUZ-2023-0019.
\end{acknowledgements}


\bibliographystyle{aa}
\bibliography{HMYSO}

\begin{appendix}


\section{Evolution of H$_2$O}\label{h2o_evo}

\begin{figure}[!ht]
    \centering
    \includegraphics[width=1.0\hsize]{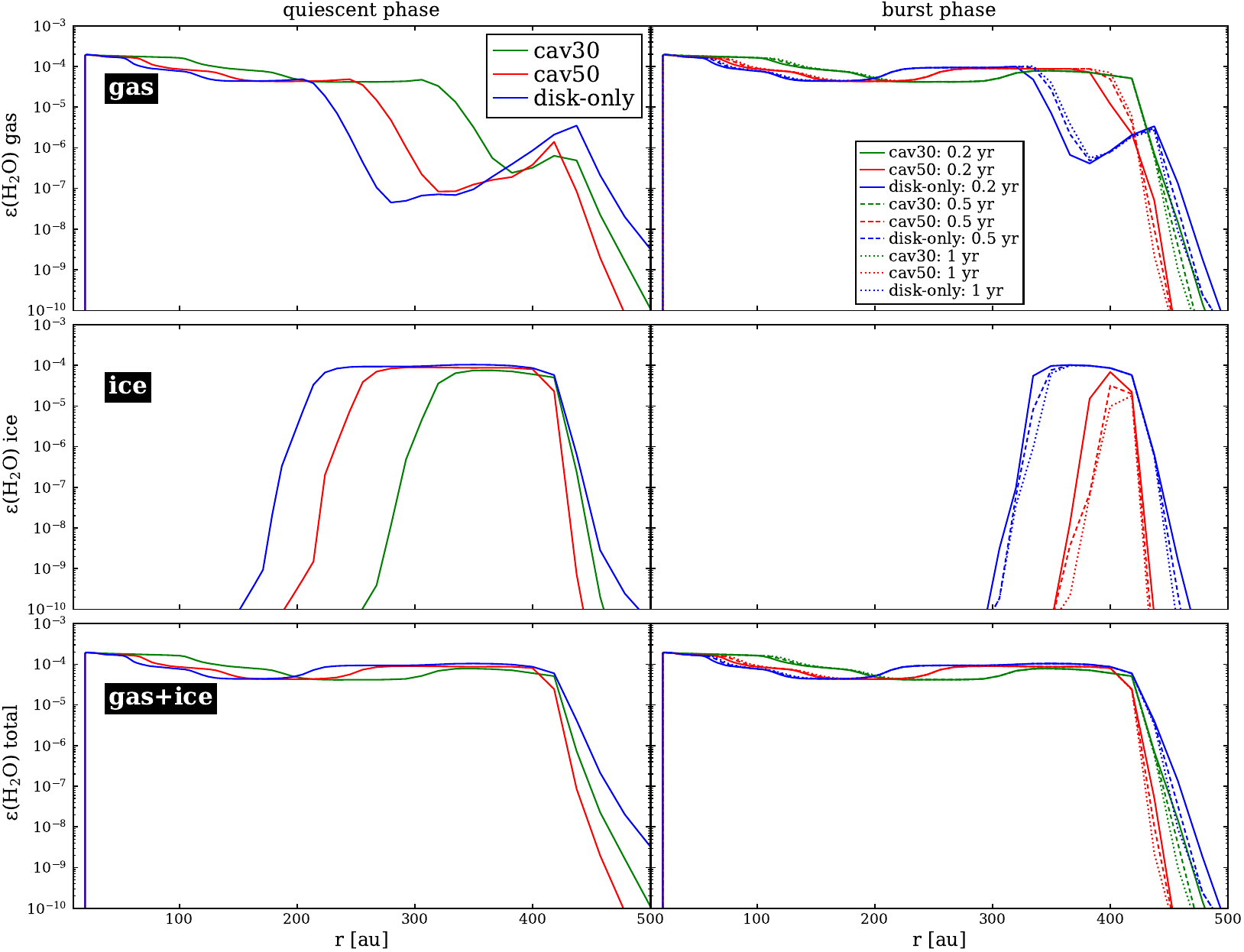}
    \caption{Averaged abundances of water for models cav30 (green), cav50 (red), and disk-only (blue). The left and right columns show the pre-burst phase and the burst phase respectively. Three different time steps during the burst are displayed. The top, middle, and bottom rows show the averaged gas, ice, and total abundances respectively.} 
    \label{fig:all_all_h2o}
\end{figure}

\begin{figure}[!ht]
    \centering
    \includegraphics[width=1.0\hsize]{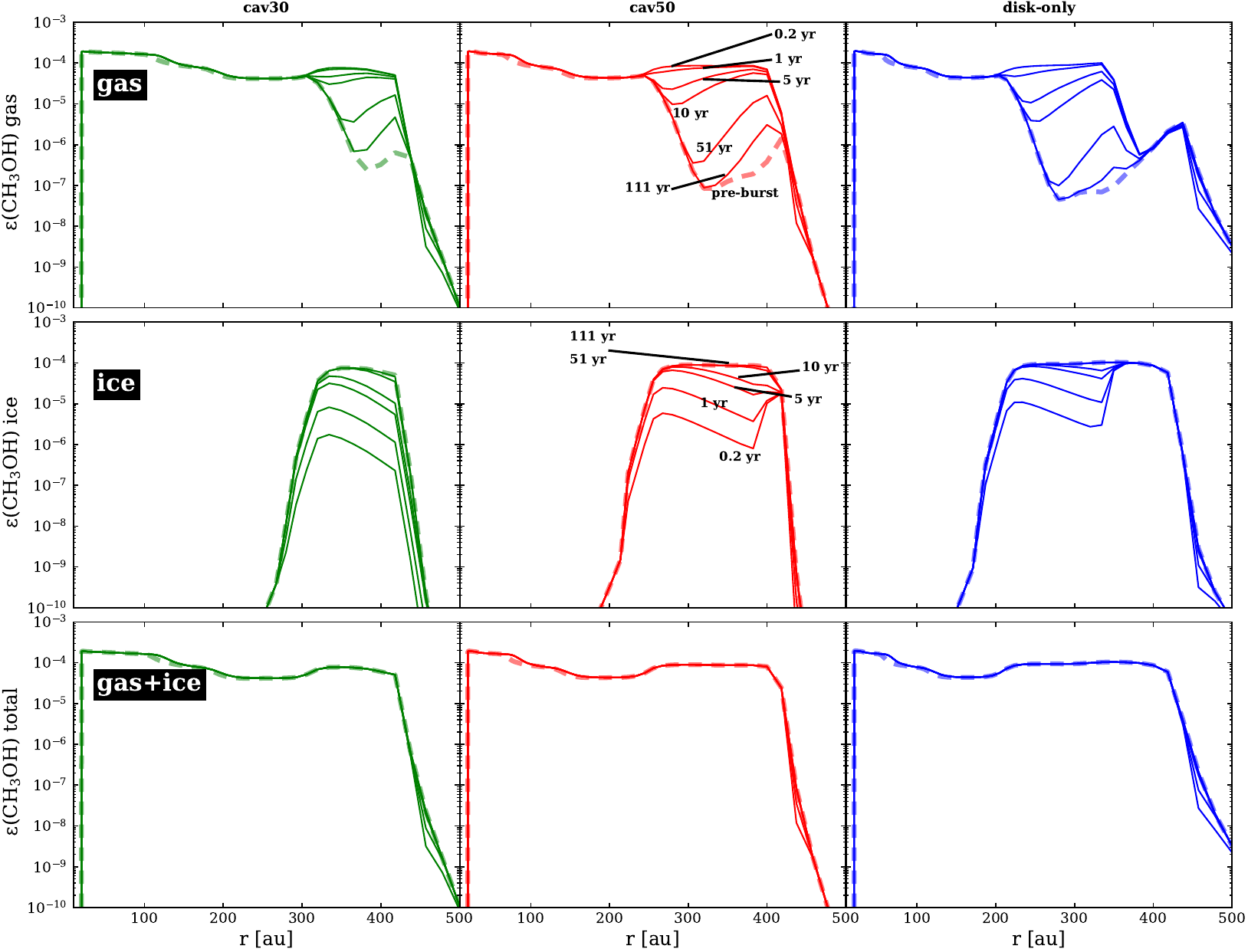}
    \caption{ {Averaged abundances of water for models cav30 (green, left column), cav50 (red, middle column), and disk-only (blue, right column) during the post-burst phase. Six different time steps during the post-burst are displayed. The top, middle, and bottom rows show the averaged gas, ice, and total abundances respectively.}} 
    \label{fig:all_pb_h2o}
\end{figure}

There are a couple of differences between the behavior of methanol and water.  {In the case of water, there seems to be a depletion of the total amount of water in the very outer parts of the disk.} This takes place around a radius of 500 au immediately after the burst (right column of Fig. \ref{fig:all_all_h2o}). Both gas and water ice show the same trend for that region of the object. This depletion is due to dissociation processes caused by a stronger radiation field during the accretion burst. For example, the main destruction reaction for gaseous water in that region of the object is the photodissociation of H$_2$O with OH and H as products.

Furthermore, there is a clear spike for gas-phase water at around 450 au shown in the top-right panel of Fig. \ref{fig:all_all_h2o}. This spike is not affected by the burst but for smaller or greater radii the values for the water gas abundances do change over time. This spike represents the outer edge of a cavity in the gaseous water abundance that is filled with water ice and extends from 200 to 420 au for the pre-burst phase and shrinks during the burst. The spike is the region in the object where the water ice is desorbed by the external UV radiation field. The UV field is able to photo-desorb and photo-dissociate water ice in that region due to the low densities present there.
In Appendix \ref{2D cuts} we show the radial and vertical abundance distribution which clearly displays the mentioned cavity. The extension of this cavity during the burst is from 350 to 420 au. As the position of the outer edge of this cavity is not affected by the burst, the values for the gas and ice abundances around that radius will not show a significant response to the burst. This peak is also present in the two previous models (cav30 and cav50). However, because the region (between 400 and 500 au) where water ice and gaseous water are both present is less extended than in the disk-only model the mentioned spike is less visible.

\clearpage

\section{2D vertical cuts}
\label{2D cuts}
{In this section we present a complementary view on the abundance evolution of methanol and water that we provide in section \ref{chem_evo}. The figures in that section show the averaged abundance of the CH$_3$OH and H$_2$O. The averaged abundance as a function of the radius could dismiss local variations along the vertical abundance distribution of the disk. In order to prevent this, we show the radial and vertical abundance distribution (Figs. \ref{fig:c3_2D_evo}, \ref{fig:c5_2D_evo} and \ref{fig:d_2D_evo}) of the three models for the most significant snapshots of their chemical evolution (pre-burst, 1 year after the start of the burst, 0.2, 10, and 111 years post-burst phase). The 2D cuts shown here confirm the trends already described in section \ref{chem_evo}. }

\begin{figure}[!h]
    \begin{subfigure}[t]{0.5\textwidth}
    \includegraphics[width=\textwidth]{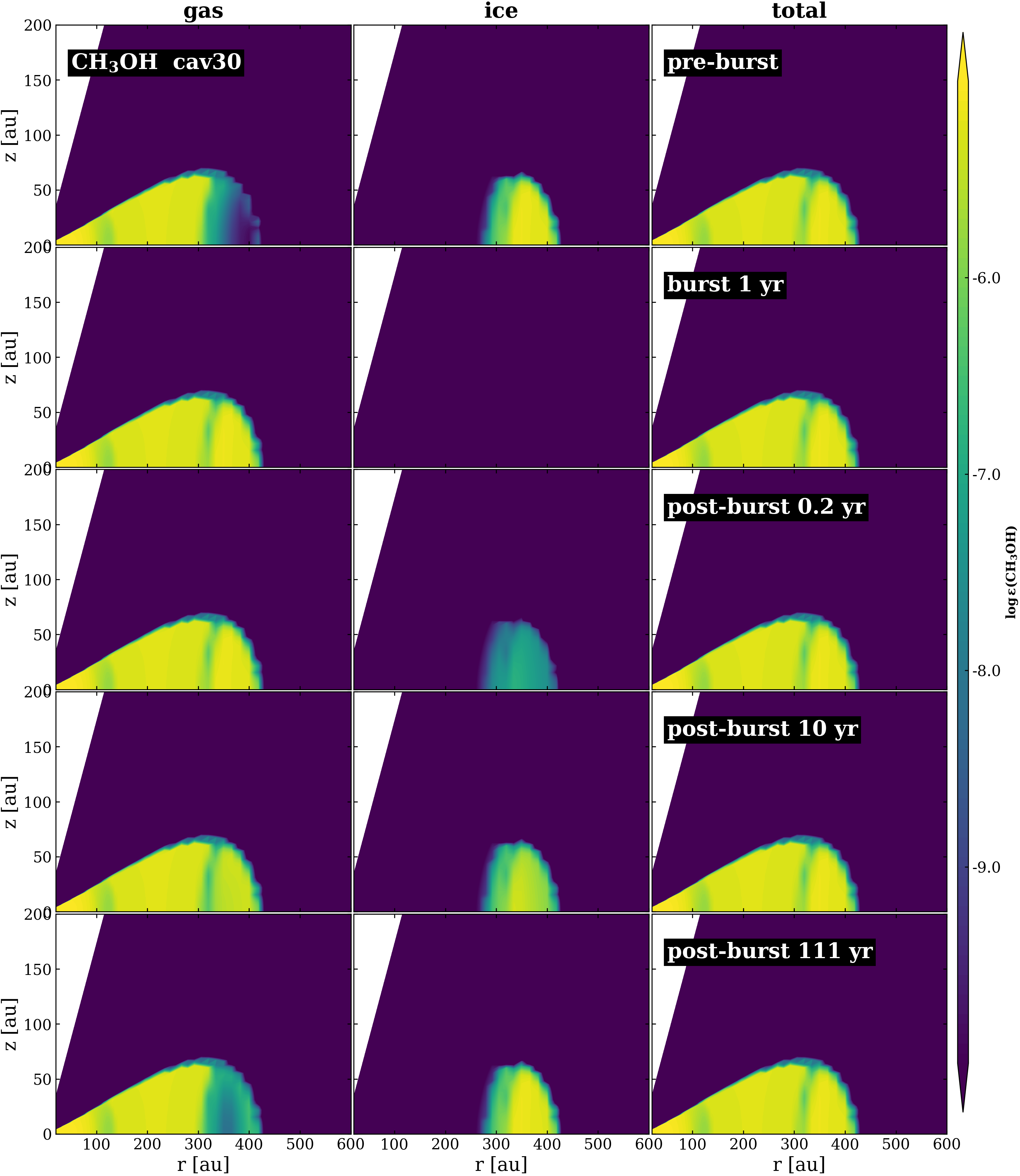}
    \end{subfigure}
    \hfill
    \begin{subfigure}[t]{0.5\textwidth}
    \includegraphics[width=\textwidth]{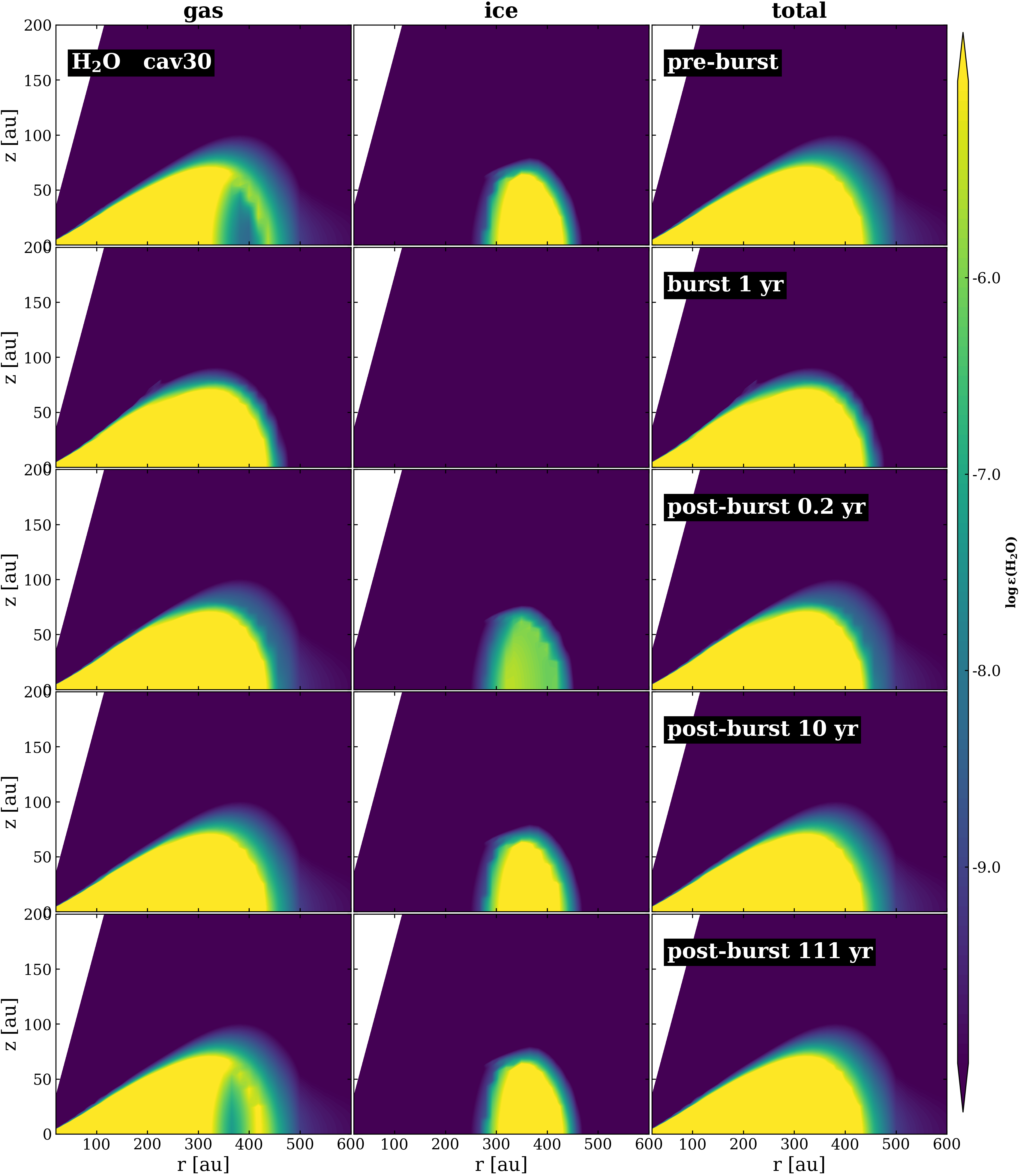}
    \end{subfigure}
    \caption{Snapshots of the radial and vertical abundance distribution of CH$_3$OH (top) and H$_2$O (bottom) of the cav30 model.}
    \label{fig:c3_2D_evo}
\end{figure}

\begin{figure}[!h]
    \begin{subfigure}[t]{0.5\textwidth}
    \includegraphics[width=\textwidth]{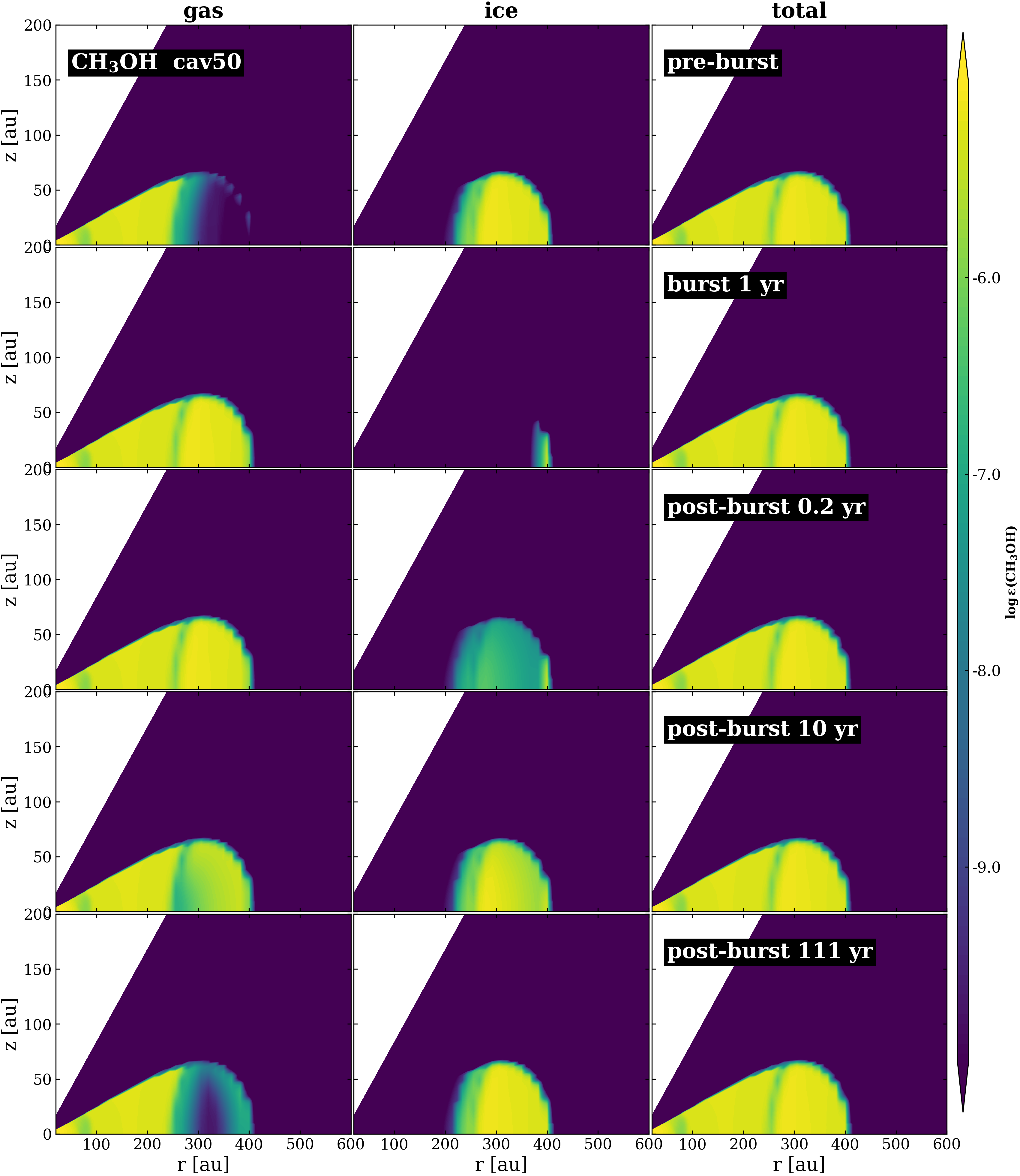}
    \end{subfigure}
    \hfill
    \begin{subfigure}[t]{0.5\textwidth}
    \includegraphics[width=\textwidth]{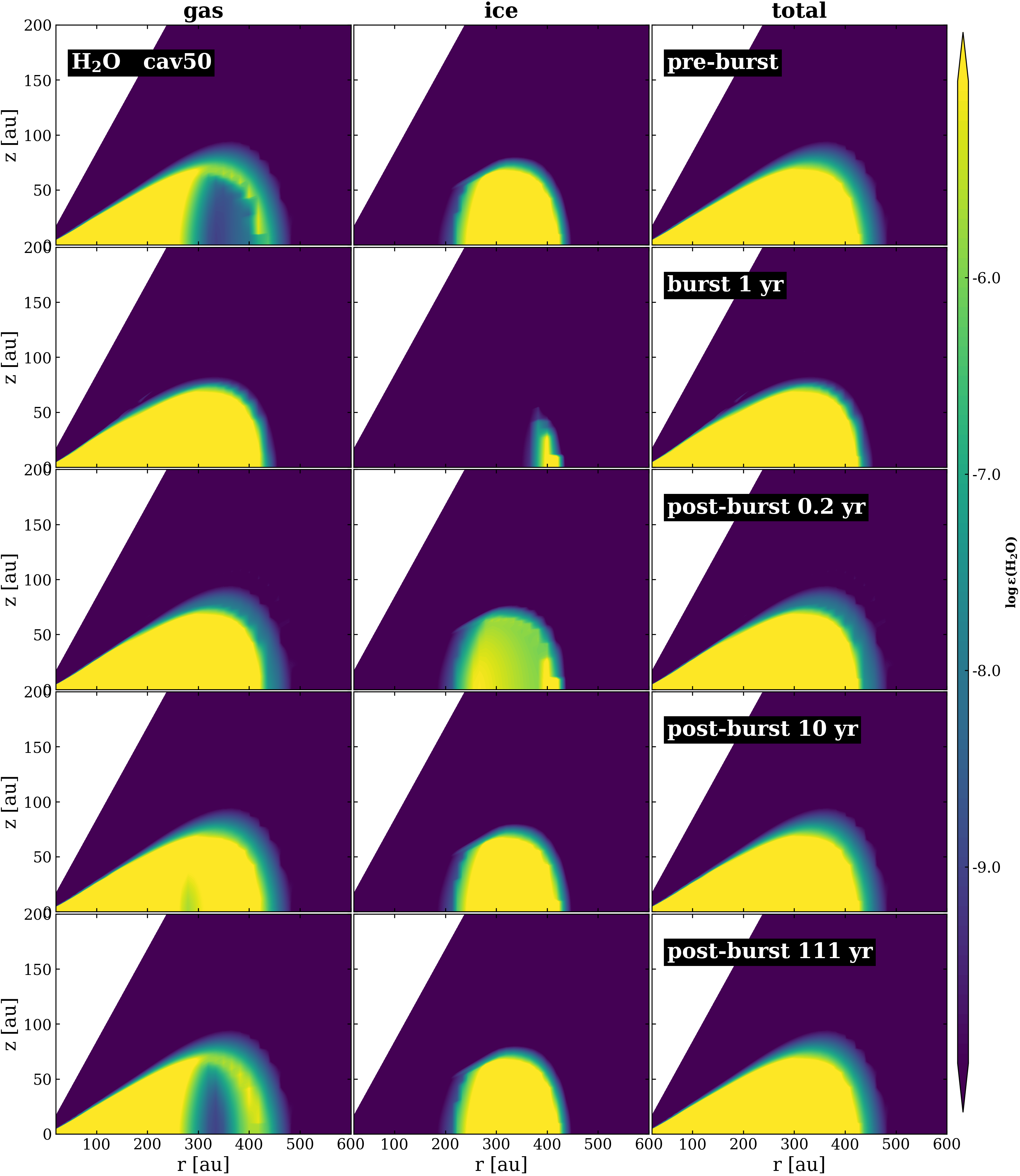}
    \end{subfigure}
    \caption{Snapshots of the radial and vertical abundance distribution of CH$_3$OH (top) and H$_2$O (bottom) of the cav50 model.}
    \label{fig:c5_2D_evo}
\end{figure}

\begin{figure}[!h]
    \begin{subfigure}[t]{0.5\textwidth}
    \includegraphics[width=\textwidth]{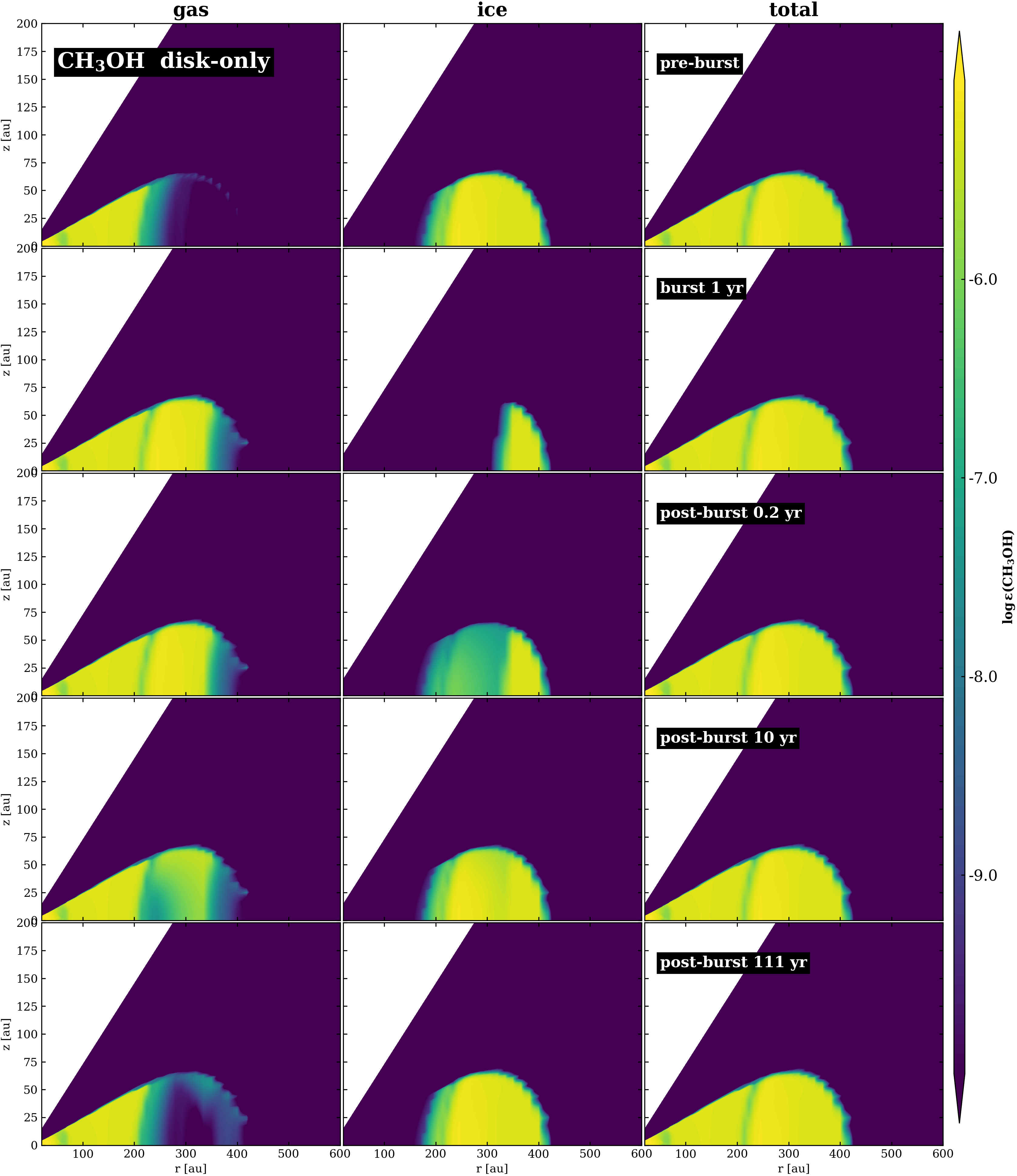}
    \end{subfigure}
    \hfill
    \begin{subfigure}[t]{0.5\textwidth}
    \includegraphics[width=\textwidth]{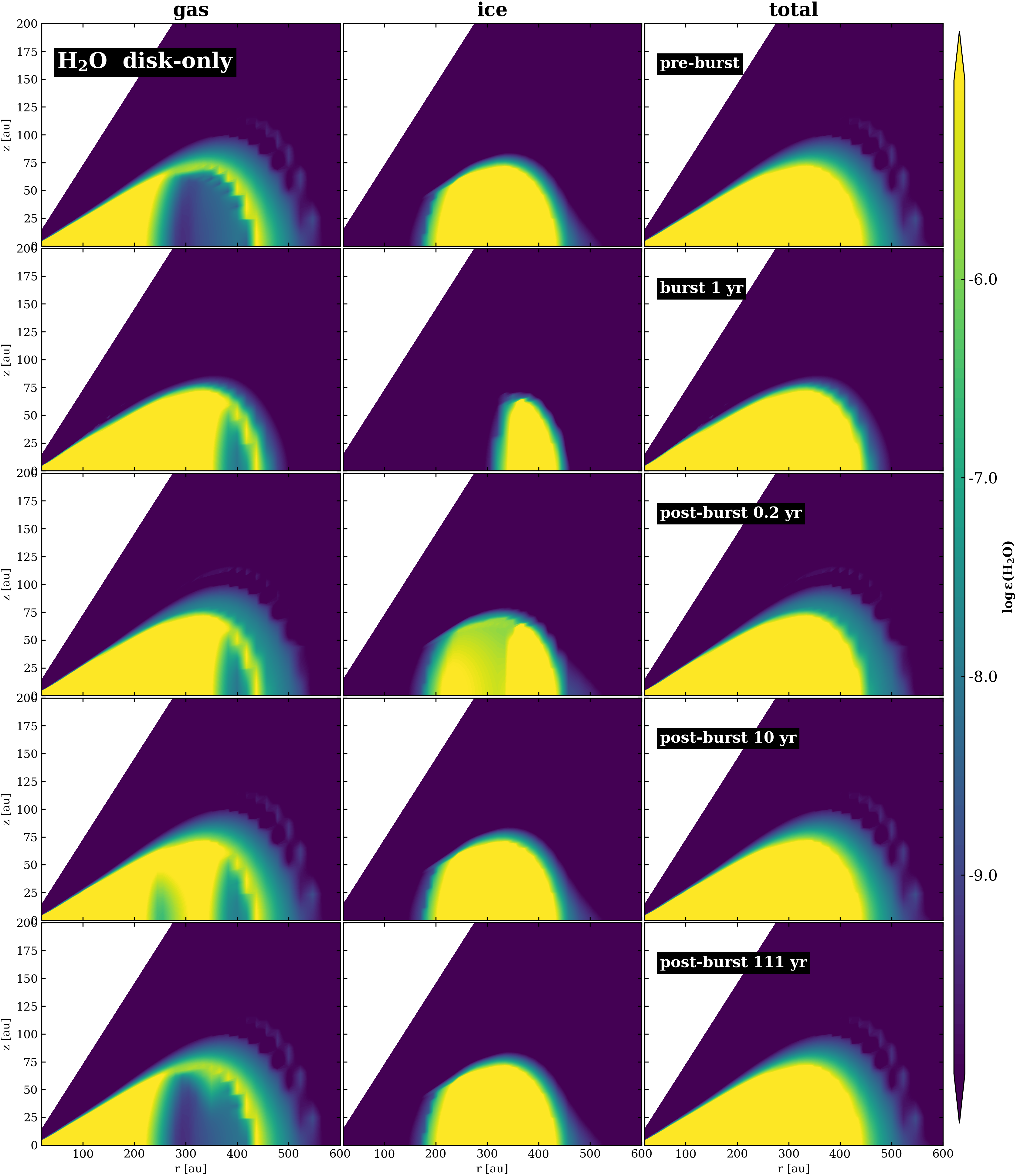}
    \end{subfigure}
    \caption{Snapshots of the radial and vertical abundance distribution of CH$_3$OH (top) and H$_2$O (bottom) of the disk-only model.}
    \label{fig:d_2D_evo}
\end{figure}

\clearpage

\section{Longer lasting bursts H$_2$O}
\label{longer busts}

 The results presented here are the same as those presented in Fig. \ref{fig:t_ice_gas_ratio} but for the water molecule. Here the ice-to-gas ratio behaves similarly to the one shown for methanol. Most desorption happens during the first three years of the burst. After the burst the the quiescent state is reached also after approximately 100 years.
 
 \begin{figure}[!h]
  \centering
  \begin{subfigure}[t]{0.9\hsize}
\includegraphics[width=\hsize]{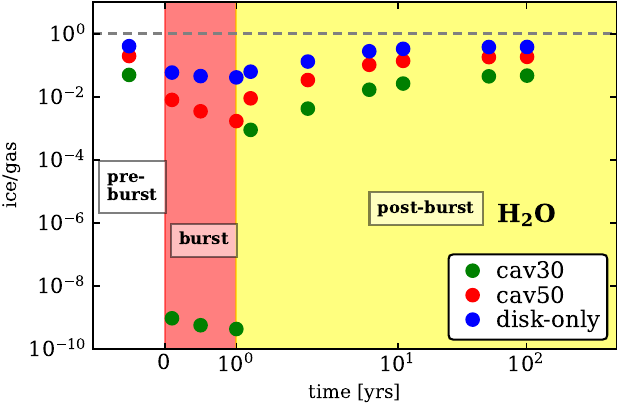}
    \end{subfigure}
    \begin{subfigure}[t]{0.9\hsize}
\includegraphics[width=\hsize]{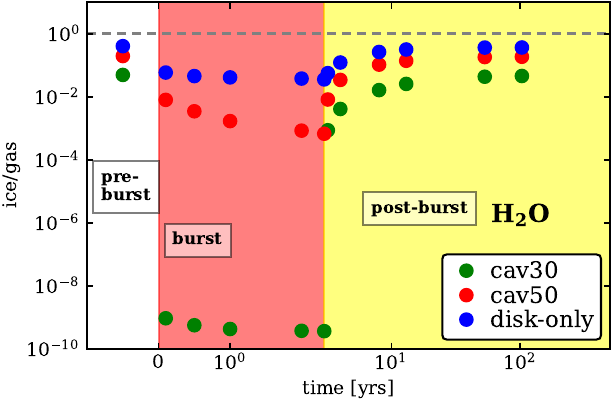}
    \end{subfigure}
\begin{subfigure}[t]{0.9\hsize}
\includegraphics[width=\hsize]{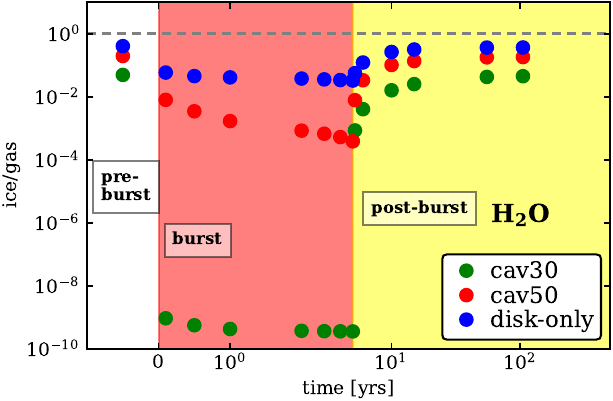}
    \end{subfigure}
\caption{Ice-to-gas ratio evolution over time. The white, pink,
 and yellow regions represent the pre-burst, burst, and post-burst phases respectively.
The green (model cav30), red (model cav50), and blue (model disk-only) dots show the ice-to-gas ratio as a function of time.
 The top, middle, and bottom panels show the ice-to-gas ratio of H$_2$O for the setups discussed so far for a burst
  duration of 1, 3, and 5 years, respectively.}
\label{fig:longer_burst}
\end{figure}


\end{appendix}

\end{document}